\documentclass[pdflatex,sn-basic]{sn-jnl}

\usepackage{graphicx}
\usepackage{booktabs}
\usepackage{amsmath}
\usepackage{amssymb}
\usepackage{xcolor}
\usepackage[shortlabels]{enumitem}
\usepackage{multirow}
\usepackage{tabularx}
\usepackage{float}
\usepackage{subcaption}
\usepackage{algorithm}
\usepackage{algpseudocode}
\usepackage{url}
\usepackage{caption}
\usepackage{hyperref}
\hypersetup{colorlinks=true,linkcolor=blue,citecolor=blue,urlcolor=blue}

\newcommand{\system}{\textsc{Text2NGAC}}
\newcommand{\ie}{\textit{i.e.}}
\newcommand{\eg}{\textit{e.g.}}
\newcommand{\etal}{\textit{et~al.}}

\providecommand{\FloatBarrier}{}

\begin{document}

\title[Text2NGAC: multi-LLM NGAC policy extraction]{\system{}: a multi-LLM pipeline for NGAC policy extraction with mutation-based verification}

\author[1]{\fnm{Ayesha} \sur{Habib}}\email{ayesha.habib@mail.nwpu.edu.cn}

\author*[1]{\fnm{Huixiang} \sur{Zhang}}\email{zhanghuixiang@nwpu.edu.cn}

\author[1]{\fnm{Dan} \sur{Zhu}}\email{zhudan@nwpu.edu.cn}

\author[2]{\fnm{Waiz} \sur{Haider}}\email{waiz.haider@mail.nwpu.edu.cn}

\affil*[1]{\orgdiv{School of Cybersecurity}, \orgname{Northwestern Polytechnical University}, \orgaddress{\city{Xi'an}, \postcode{710129}, \country{China}}}

\affil[2]{\orgdiv{School of Computer Science}, \orgname{Northwestern Polytechnical University}, \orgaddress{\city{Xi'an}, \postcode{710129}, \country{China}}}

\abstract{Access-control policies are written in natural language but enforced as formal models, and translating them into NIST Next Generation Access Control (NGAC) graphs remains unsolved: prior systems cover only subsets of the NGAC model, and none verify the graphs they extract. We present \system{}, a pipeline of cooperating large language models that disambiguates each sentence before extraction, escalates uncertain sentences to a multi-model ensemble, and checks the resulting graph with a five-strategy verifier whose own effectiveness is measured by controlled mutation testing. To our knowledge it is the first system to extract all four NGAC relationship types --- associations, prohibitions, event--response obligations, and assignments --- from real-world policy documents. Three results anchor the evaluation. The verifier detects 99.7\% of 918 injected mutants and every structural fail-open fault, and the fault classes it misses are characterized and weighted by security consequence rather than left unexamined. The extracted graphs are enforceable: one corpus's graph loads unmodified into the NIST reference implementation, which reproduces all 55 access decisions the graph entails and fires an extracted obligation end-to-end. Extraction reaches 97.3--99.6\% exact-tuple-match accuracy on the 714 scored sentences of four real corpora, far above three reproduced baselines on matched metrics, and paraphrase probes with semantic re-scoring bound the pipeline's sensitivity to rewording. Reported accuracy is wording- and benchmark-specific, adversarial input is outside the verifier's guarantees, and the intended deployment is extraction plus structural verification feeding expert review. A complete two-tier reproducibility artifact accompanies the paper.}


\maketitle

\section{Introduction}
\label{sec:introduction}

Access control policies (ACPs) are vital to protect resources against unauthorized access while guaranteeing accessibility to legitimate users. ACPs are typically part of security requirements, often written in natural language. They articulate strict rules describing how access is managed, who may access resources, and under what conditions. ACP statements are manually translated into a formal representation (\ie, a security model)~\citep{hu2013guide} following a standard framework such as Role-Based Access Control (RBAC)~\citep{sandhu1996role,ferraiolo2001proposed} or Attribute-Based Access Control (ABAC). Next Generation Access Control (NGAC), standardized in NIST SP 800-178~\citep{ferraiolo2016ngac}, unifies attribute-based, role-based, and discretionary access control through a graph-based architecture with four relation types (associations, prohibitions, obligations, and assignments) and enforcement via the NIST Policy Machine~\citep{ferraiolo2015pm}.

Researchers have worked on automated extraction of ACP elements using NLP and machine learning techniques~\citep{narouei2018identification,xiao2012automated,slankas2014relation,alohaly2019deep}. However, existing approaches to NGAC policy extraction face four critical challenges that remain unresolved.

\noindent\textbf{Challenge~1: Incomplete NGAC coverage.} Prior work covers only subsets of the NGAC model: rule-based extraction reaches associations and partial assignments at ${\sim}75\%$ accuracy~\citep{abdelgawad2023synthesizing}; LLM prompting reports 99\% F1, but only on LLM-synthesized data and without obligations or verification~\citep{lawal2024translating}; a NIST report excludes assignments entirely~\citep{nist2024translating}; and ABAC-oriented work targets flat Subject--Action--Resource triples~\citep{yang2025extraction}. No existing approach supports all four NGAC types with obligations modeled as event--response structures per NIST SP 800-178.

\noindent\textbf{Challenge~2: Linguistic complexity.} Natural language policy documents contain compound sentences, pronoun ambiguity, and implicit subjects that confound single-pass extraction. For example, ``Only the name and email are provided by the HCP'' contains a conjunction requiring decomposition and a passive voice requiring subject resolution. The syntactic parsing of \citet{abdelgawad2023synthesizing} produces systematic failures on such constructs. Single-model approaches also exhibit per-type biases: our evaluation of 15 LLMs reveals that no single model dominates all NGAC types, with Assignment accuracy varying from 54\% to 100\% across models.

\noindent\textbf{Challenge~3: Verification gap.} Extracted policy correctness must be validated against formal NGAC semantics, yet \emph{zero} of the four most relevant prior NL-to-NGAC systems~\citep{abdelgawad2023synthesizing,lawal2024translating,yang2025extraction,nist2024translating} provide automated verification with measured effectiveness: \citet{abdelgawad2023synthesizing} include only partial rule-based checking, while \citet{lawal2024translating}, \citet{yang2025extraction}, and the NIST report~\citep{nist2024translating} include no automated verification at all. Manual verification does not scale, and a single misclassified relationship in the 725-sentence corpus of this paper would suffice to silently create a security vulnerability.

\noindent\textbf{Challenge~4: Verification rigor.} Even when a verification phase is included, the question of whether it is \emph{effective} at catching real errors is rarely answered. A pipeline that passes all of its own verification checks could still be wrong if the checks are too lenient. Across the same four prior systems above, \emph{none} report quantitative evidence, such as detection rate against controlled fault injection, that the verifier (where present) would catch a deliberately corrupted policy graph.

\noindent\textbf{Our contributions.} In response to the above four challenges, we develop three integrated innovations:

\begin{enumerate}[label=\textbf{(\arabic*)},leftmargin=2.2em]
 \item \textbf{Proactive semantic disambiguation (Phase~0).} We decouple linguistic complexity from extraction by resolving ambiguity before tuple extraction begins. An LLM-driven scoring rubric flags compound sentences, passive voice, pronoun ambiguity, and implicit subjects; flagged sentences are decomposed and rewritten by a comprehension--rewrite cycle, with an independent LLM validating that semantics are preserved. This addresses Challenge~2; its measured contribution is snapshot-dependent --- largest where the base model is weakest (Section~\ref{sec:ablation}).

 \item \textbf{Task-optimized multi-LLM ensemble (Phase~A).} A six-stage extraction pipeline routes each sentence through policy-class assignment, RAG-augmented type classification, and tuple extraction. A two-valued confidence-gated escalation mechanism (Section~\ref{sec:threshold-sensitivity}) routes low-confidence sentences to an ensemble that exploits empirically-measured per-type model complementarity. This addresses Challenges~1 and~2.

 \item \textbf{Five-strategy verification paired with controlled-mutation evaluation (Phase~B).} Phase~B verifies the extracted graph against the original sentences through five complementary strategies: coverage (every sentence is reflected), correctness (entities are well-typed), consistency (no allow/deny conflicts), obligation completeness (every obligation has both event and response), and minimality (no redundant edges). To assess the verifier's effectiveness against extractor errors, we adapt the NGAC mutation methodology of \citet{chen2021mutation} to the NL-extracted-graph setting: eight controlled fault-injection operators corrupt the verified graph in well-defined ways, and we measure the fraction of corruptions that Phase~B detects, classifying every injected fault by its security consequence: fail-open faults that silently broaden privilege versus fail-closed faults that merely deny legitimate access. Under this classification, every structural fail-open fault is detected (438/438), while semantically plausible privilege-broadening faults are detected at only 37.7\% --- a measured boundary (Section~\ref{sec:severity}). The novelty is narrower than Chen \etal's ``mutation testing for NGAC'': we apply mutation analysis to validate a verifier of \emph{NL-extracted} policies (Section~\ref{sec:related}), addressing Challenges~3 and~4.
\end{enumerate}

The paper proceeds from preliminaries and datasets (Section~\ref{sec:preliminaries}) through the three-phase pipeline (Section~\ref{sec:approach}), the extraction results (Section~\ref{sec:evaluation}), the controlled-mutation evaluation, the prompt-injection analysis, and the enforcement replay (Sections~\ref{sec:rigor}--\ref{sec:pm-replay}), related work (Section~\ref{sec:related}), and limitations (Section~\ref{sec:discussion}).

\FloatBarrier
\section{Preliminaries}
\label{sec:preliminaries}

\subsection{NGAC}
\label{sec:ngac}

The NGAC framework~\citep{ferraiolo2016ngac} models access control as a directed graph in which each \emph{user} (a principal) is grouped into one or more \emph{user attributes} (think roles, departments, clearances), each \emph{resource} (a protected object) is grouped into one or more \emph{resource attributes} (think folders, file types, sensitivity classes), and access is granted or denied through edges that connect those attributes. The graph carries four relationship types: an \emph{association} edge says ``users in attribute~$X$ may perform operation~$o$ on resources in attribute~$Y$,'' a \emph{prohibition} edge says the opposite (and wins under deny-overrides), an \emph{obligation} edge declares an event-triggered automatic response (``when the registration period ends, lock the form''), and an \emph{assignment} edge places a child entity inside a parent attribute (\eg, placing role \texttt{LHCP} inside role \texttt{HCP}, or placing concrete record \texttt{pat\_001} inside attribute \texttt{patient\_record}). A \emph{policy class} is an administrative container that bounds the scope of a hierarchy. Access decisions are computed by traversing this graph; a request \texttt{(user, op, resource)} is allowed iff there exists a path through user attributes and resource attributes whose final association covers~\texttt{op} and no prohibition along that path forbids it. The remainder of this section formalizes the entities and relationships described above.

\smallskip\noindent\textbf{Definition 1 (Entity types).} NGAC distinguishes six vertex types: users (U), user attributes (UA), resources (R, also called \emph{objects} in the NGAC literature), resource attributes (RA), policy classes (PC), and operations. Users represent identifiable principals; user attributes group users by role, clearance, or department. Resources are protected items; resource attributes group them by sensitivity, type, or owner. Policy classes are administrative containers that bound the scope of attribute hierarchies.

\smallskip\noindent\textbf{Definition 2 (Association).} An access permission: \texttt{assoc}(\textit{ua}, \textit{op}, \textit{ra}), granting users in user attribute \textit{ua} the right to perform operation \textit{op} on resources in resource attribute \textit{ra}.

\smallskip\noindent\textbf{Definition 3 (Prohibition).} An access denial: \texttt{prohib}(\textit{ua}, \textit{op}, \textit{ra}), denying users in \textit{ua} the right to perform \textit{op} on resources in \textit{ra}. Prohibitions take precedence over associations under NGAC's deny-overrides semantics.

\smallskip\noindent\textbf{Definition 4 (Obligation).} An event-triggered automated response: \texttt{oblig}(\textit{event}, \textit{response}). True obligations require \emph{both} an EVENT trigger describing an observable action \emph{and} an automatic RESPONSE specifying the system's mandatory reaction. Simple workflow descriptions or conditional permissions are not obligations per NIST SP 800-178~\citep{ferraiolo2016ngac}.

\smallskip\noindent\textbf{Definition 5 (Assignment).} Hierarchical containment: \texttt{assign}(\textit{child}, \textit{parent}). NGAC supports six sub-types: U$\to$UA, UA$\to$UA, R$\to$RA, RA$\to$RA, UA$\to$PC, and RA$\to$PC. \footnote{We write \emph{resources} where SP 800-178 writes \emph{objects} (Definition~1); our R/RA sub-types map one-to-one onto the standard's O$\to$OA, OA$\to$OA, and OA$\to$PC vocabulary, so a reader checking conformance can substitute O for R and OA for RA throughout.} \citet{tan2025safety} proved the NGAC safety problem coNP-complete under mild assumptions, underscoring the need for automated policy construction to ensure correctness at scale.

\begin{figure}[!tbp]
 \centering
 \includegraphics[width=0.65\columnwidth]{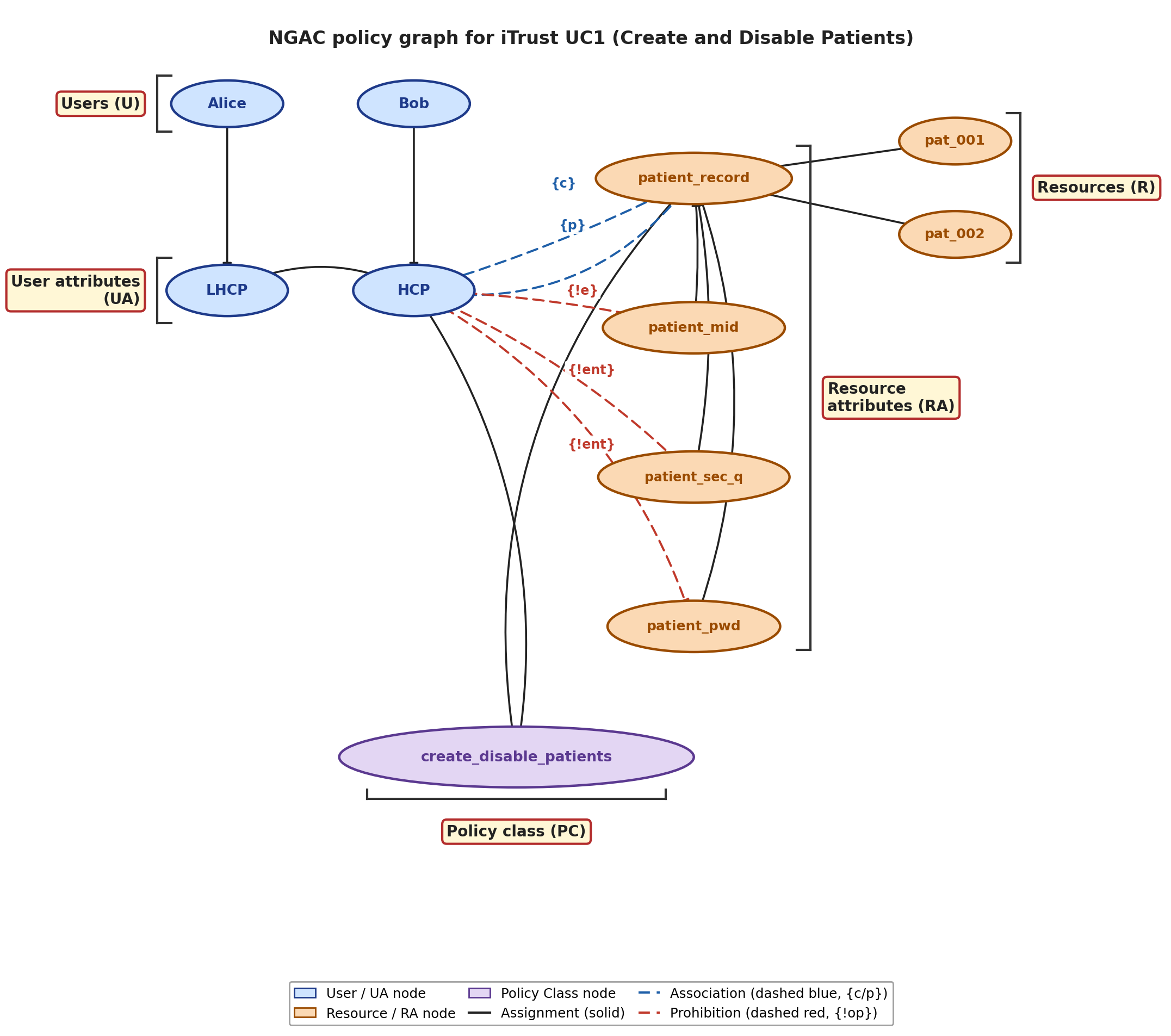}
 \caption{NGAC policy graph for the running example (iTrust UC1). Node and edge color codings are given by the in-figure key (\texttt{!e} = deny edit, \texttt{!ent} = deny enter). A simplified subset; the full released Stage-6 graph appears in Figure~\ref{fig:uc1_graph}.}
 \label{fig:ngac_example}

\end{figure}

Figure~\ref{fig:ngac_example} illustrates a small NGAC policy graph with five of the six entity types of Definition~1 --- operations label edges rather than appearing as vertices --- and three relationship types.

The graph concretely encodes the four iTrust UC1 sentences used as the running example throughout Section~\ref{sec:approach} (UC1-2, UC1-4, UC1-7, UC1-8; the prefix avoids collision with Phase~B's S1--S5 strategy labels): each labeled edge in the figure derives from one of these sentences, traced stage by stage in Table~\ref{tab:walkthrough}, and the UA/RA hierarchies are extracted in Stage~5 from possessives such as ``the patient's MID''. All edges are anchored to the policy class \texttt{create\_disable\_patients} per the UC1 header.

\subsection{Design-Time Model Evaluation}
\label{sec:design-time}

Before constructing the pipeline, we benchmarked 15 LLMs across the four NGAC relationship types on a balanced 50-sentence design-time set (10 sentences from each of five sources; no design-time sentence appears in any evaluation set; protocol, footnoted caveats, and the full per-model heat map ship with the artifact). Two findings drive the architecture: no single LLM dominates all four types (per-type accuracy 54\%--100\%, with different models excelling on different types), and the complementarity is exploitable by a confidence-weighted vote. With $n{=}11$--$14$ sentences per type cell these are motivating signal, not load-bearing evidence. The decisive evidence for the ensemble comes from the evaluation corpora at natural scale (Sections~\ref{sec:eval_stage3} and~\ref{sec:ablation}).

\subsection{Datasets}
\label{sec:datasets}

We evaluate on four datasets (Table~\ref{tab:datasets}), representing the most widely used benchmarks in NL-to-access-control research~\citep{lawal2024translating}.

\noindent\textbf{iTrust Text2Policy}~\citep{meneely2012itrust,xiao2012automated} is a healthcare EHR system (234~ACP sentences across 37 use cases, 36 of which define policy classes; 12 prohibitions on sensitive patient data), the standard iTrust corpus of Xiao \etal's Text2Policy work. \textbf{iTrust ACRE-n}~\citep{slankas2014relation,meneely2012itrust} is the larger iTrust corpus (362~ACPs) from the ACRE relation-extraction release, with more explicit-trigger obligations and substantially more compound sentences --- a same-domain stress test that exposes failure modes invisible on the smaller variant. \textbf{CyberChair} is a conference-management system (54~ACPs) with complex assignment hierarchies and multi-level role inheritance. \textbf{IBM Course Registration}~\citep{ibm2010course} is a university system (75~ACPs) heavily skewed toward associations (85.3\%), stressing the type-distribution prior on which Stage~3 R1 relies.

Our combined corpus of 725~ACP sentences across four independently-authored documents from three domains is comparable to or exceeds recent related work: \citet{abdelgawad2023synthesizing} evaluate on $\sim$200 sentences; \citet{lawal2024translating} on $\sim$250 LLM-synthesized sentences. Publicly available NL-to-access-control datasets remain scarce: unlike NER or relation extraction, no large-scale NGAC-annotated corpus exists, and constructing one requires expertise in both access control semantics and the NIST SP 800-178 type system.

\begin{table*}[!tbp]
\centering
\caption{Dataset characteristics.}
\label{tab:datasets}
\small
\begin{tabular}{@{}lrrrrrrr@{}}
\toprule
\textbf{Dataset} & \textbf{Tot.} & \textbf{ACP} & \textbf{Asc} & \textbf{Asg} & \textbf{Obl} & \textbf{Prh} & \textbf{P0} \\
\midrule
iTrust Text2Policy & 471 & 234 & 168 & 45 & 9 & 12 & 67 \\
iTrust ACRE-n & 1160 & 362 & 238 & 81 & 36 & 7 & 143 \\
CyberChair & 303 & 54 & 29 & 19 & 6 & 0 & 21 \\
IBM Course Reg. & 401 & 75 & 64 & 6 & 1 & 4 & 14 \\
\midrule
\textbf{Total} & 2335 & 725 & 499 & 151 & 52 & 23 & 245 \\
\bottomrule
\end{tabular}
\par\smallskip{\raggedright\footnotesize Tot.~= total sentences in source document; ACP = access-control sentences; Asc/Asg/Obl/Prh = relationship-type counts; P0 = sentences modified by Phase~0.\par}
\end{table*}

\smallskip\noindent\textbf{Ground truth construction.} Annotations were created by the authors following the NGAC relationship-type criteria of NIST SP 800-178. Annotation comprised sentence-level type classification against the formal definitions of Section~\ref{sec:ngac} (associations require an explicit actor--operation--object triple; obligations require both an EVENT trigger and an automatic RESPONSE; assignments describe hierarchical containment), tuple extraction under strict NGAC naming conventions, and cross-validation against the original policy documents. Ground-truth reliability was assessed by an independent second annotator outside the annotation lineage, who coded all four corpora against these NGAC type criteria. Cohen's $\kappa$ against the released labels is \textbf{0.84 pooled} ($n{=}724$; one iTrust Text2Policy sentence the coder left blank is excluded): 0.93 on iTrust Text2Policy, 0.82 on iTrust ACRE-n, 0.72 on IBM, and 0.70 on CyberChair; both codings are released with the artifact. Per-type (one-vs-rest) $\kappa$ on the same codings: association 0.70--0.96, assignment 0.58--0.92, prohibition 0.87--1.00 where present, and obligation 0.21--0.79 --- weakest on CyberChair, whose ground truth holds only six obligations; CyberChair and IBM disagreements additionally went through a written-feedback adjudication round, released alongside the codings. 

\FloatBarrier
\section{Approach}
\label{sec:approach}

\noindent\textbf{Research questions.} The three contributions of Section~\ref{sec:introduction} map directly to three research questions, one per innovation phrase in the title:
\begin{itemize}[leftmargin=2.2em]
 \item \textbf{RQ1 (Automated Extraction).} Can a multi-stage LLM-driven pipeline achieve end-to-end extraction of NGAC policy graphs from real-world natural-language policy documents at accuracy levels that support an expert-in-the-loop workflow (extraction plus structural validation feeding human review, Section~\ref{sec:limitations}), rather than unsupervised deployment? \emph{Answer:} \textbf{97.3--99.6\%} sentence-level exact-tuple-match accuracy on the corpora's original wording, far above all three reproduced baselines on matched metrics (Sections~\ref{sec:eval_stage4}, \ref{sec:baseline}, \ref{sec:paraphrase}).
 \item \textbf{RQ2 (Multi-LLM Pipeline).} Does confidence-gated multi-LLM ensemble routing (the gate is R1's self-reported confidence, binary in practice; Section~\ref{sec:threshold-sensitivity}) improve extraction accuracy over single-LLM configurations, and what is the marginal contribution of each pipeline component? \emph{Answer:} the ensemble-plus-refinement layer is the dominant, snapshot-stable contributor (\textbf{+13.3 to +18.8~pp} over single-model extraction); no single frontier model from three vendors (GPT-5, Claude-Sonnet-5, DeepSeek-Reasoner) closes the gap, and Phase-0/RAG contributions are snapshot-dependent (Section~\ref{sec:ablation}).
 \item \textbf{RQ3 (Mutation-Based Verification Rigor).} Is Phase~B's five-strategy automated verifier empirically effective at detecting controlled corruptions of the extracted graph, and does this effectiveness generalize across datasets? \emph{Answer:} \textbf{99.7\%} of 918 structural mutants and 96.7--99.2\% on three of four independently sampled fault classes, with parent-entity abstraction (9.2\%) as the measured boundary (Section~\ref{sec:rigor}).
\end{itemize}

\system{} is organized into three phases: \textbf{Phase~0} performs sentence-level preprocessing (semantic disambiguation), \textbf{Phase~A} performs the core information extraction in six numbered stages (Stages 1--6), and \textbf{Phase~B} performs post-extraction graph verification. We use the letter--phase nomenclature (\emph{Phase~0/A/B}) for top-level pipeline modules and the digit--stage nomenclature (\emph{Stage~1--6}) only for the internal stages of Phase~A; Phase~0 and Phase~B are not subdivided into stages. Figure~\ref{fig:phaseA} presents the high-level architecture. Throughout this section, we illustrate each stage using sentences from iTrust's \emph{Create and Disable Patients} use case (UC1) --- sentences UC1-2, UC1-4, UC1-7, and UC1-8, whose full text appears in Table~\ref{tab:walkthrough}.

\subsection{Phase 0: Proactive Semantic Disambiguation}
\label{sec:phase0}

Phase~0 realizes Contribution~1 and addresses Challenge~C2 (linguistic complexity). Prior NL-to-NGAC systems run extraction as a single pass over raw policy sentences, which fails predictably on compound sentences, passive voice, pronoun chains, and implicit subjects. Our design choice is to \emph{decouple linguistic complexity from extraction} by inserting a dedicated disambiguation stage upstream of any tuple extraction. By the time a sentence reaches Phase~A, the surface form has been normalized; downstream stages can then focus on the typing decision instead of also having to fight the surface form. The trade-off is one additional LLM call per flagged sentence (18.7--39.5\% of inputs, see below); the benefit is that downstream R1 confidence becomes informative again, not dominated by ambiguity noise.

Phase~0 resolves three categories of ambiguity \emph{before} extraction: \textbf{compound sentences} are decomposed into individual extraction targets; \textbf{pronoun references} are resolved to antecedent entities; and \textbf{implicit subjects} are inferred from context.

Each sentence is scored for ambiguity (a final score $\geq 0.3$ triggers processing). The scoring procedure has two components: (i)~an LLM (DeepSeek-Chat) assigns a base score from 0.0 to 1.0 using a structured rubric based on four linguistic features: possessive constructions, conjunctions, unclear pronouns, and complex preposition attachment; and (ii)~a deterministic regex pass adds a pattern-boost of 0.05 per detected pattern category, capped at the [0,1] interval. The threshold of 0.3 is principled: it sits at the modal boundary of the empirical score distribution; we discuss its calibration and sensitivity in Section~\ref{sec:threshold-sensitivity}. To validate the scoring heuristic against human judgment, we manually assessed all 54 CyberChair sentences: the LLM-generated scores achieved 92.6\% agreement with human ambiguity judgments (50/54 sentences matched on the binary trigger/no-trigger decision at the 0.3 threshold).

\begin{figure}[!tbp]
 \centering
 \includegraphics[width=\columnwidth]{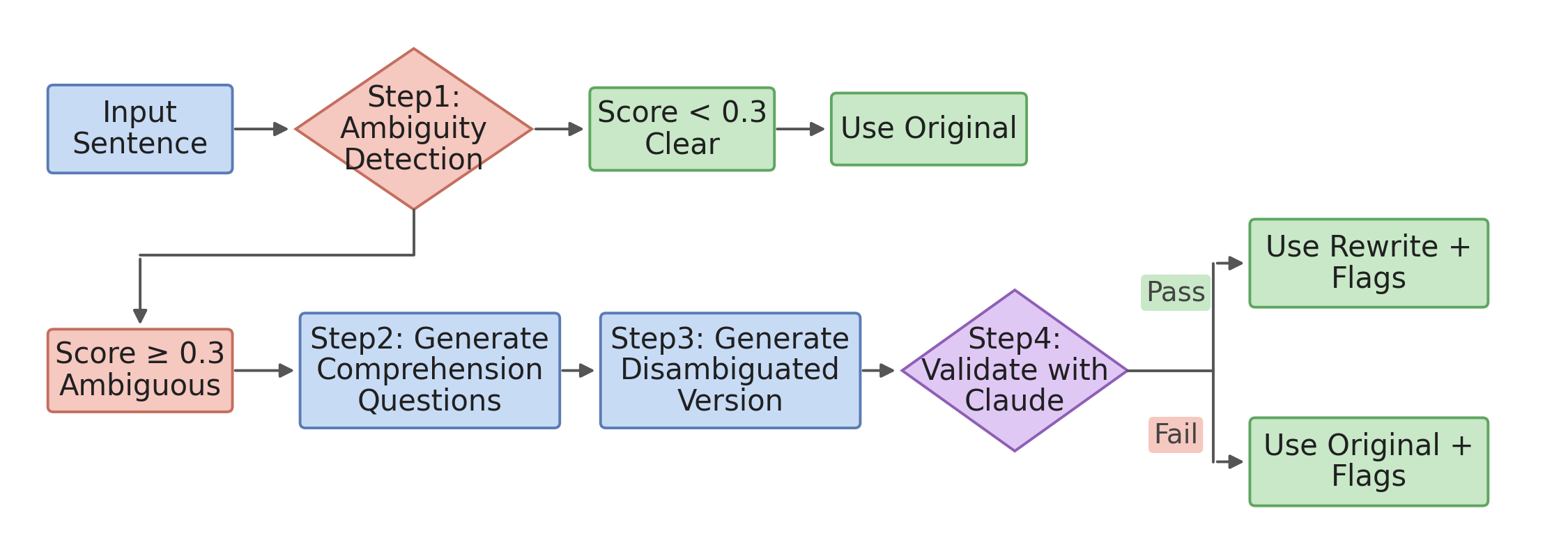}
 \caption{Phase~0 disambiguation workflow: sentences scoring $\geq$0.3 undergo comprehension-question generation, disambiguation, and independent validation; failures fall back to the original text.}
 \label{fig:phase0}

\end{figure}

When a sentence is flagged, Phase~0 executes the four-step cycle of Figure~\ref{fig:phase0}: \emph{detect} the dominant ambiguity pattern, \emph{comprehend} it via a generated question that probes the ambiguity (\eg{} ``Who provides the name and email?''), \emph{rewrite} the sentence (decomposing conjunctions, resolving passive voice, inserting inferred subjects), and \emph{validate} the rewrite with an independent LLM (Claude Sonnet, \texttt{claude-sonnet-4-5-20250929}), falling back to the original text on failure so Phase~0 cannot silently corrupt its input.

Across the four datasets, Phase~0 modified 67~of~234~iTrust Text2Policy sentences (28.6\%), 143~of~362~iTrust ACRE-n sentences (39.5\%), 21~of~54~CyberChair sentences (38.9\%), and 14~of~75~IBM sentences (18.7\%). The variation across datasets is itself informative: the ACRE-n corpus contains considerably more compound sentences and pronoun chains than the Text2Policy variant, which is part of why the same-domain comparison in Section~\ref{sec:replication} is informative.

\smallskip\noindent Table~\ref{tab:walkthrough} traces Phase~0's behavior on the running example, including its scores, the flagged issues (conjunction, passive voice, implicit subject), and the rewrites it produces.

\subsection{Phase A: Multi-Stage Extraction}
\label{sec:phaseA}

Phase~A realizes Contribution~2 and addresses Challenges~C1 (incomplete NGAC coverage) and~C2 (linguistic complexity). The two structural findings of the design-time evaluation (Section~\ref{sec:design-time}) --- no single LLM dominates all four NGAC types, and the per-type biases are exploitable by a confidence-weighted vote --- motivate Phase~A's design; the evaluation corpora confirm the pattern at natural scale, where single-model R1 trails the final pipeline by up to 33 percentage points on Stage~3 accuracy (Table~\ref{tab:stage3}). A single-LLM single-call pipeline would either inherit one model's per-type bias or pay the latency/cost of an unconditional ensemble. Our design choice is a \emph{confidence-gated ensemble}: a fast Round-1 model handles the majority of sentences alone, and only the small subset where R1 reports low confidence is escalated to a multi-LLM Round-2 ensemble. This isolates per-type bias to the sentences where it actually surfaces and keeps amortized cost low (R2 trigger rates are 3.0--20.9\% across the four corpora). The six-stage decomposition (policy class $\rightarrow$ RAG context $\rightarrow$ R1/R2 classification $\rightarrow$ tuple extraction $\rightarrow$ refinement $\rightarrow$ graph construction) further factors the problem so that each stage has a single well-defined responsibility, which makes per-stage diagnosis tractable.

\begin{figure}[!tbp]
 \centering
 \includegraphics[width=\columnwidth]{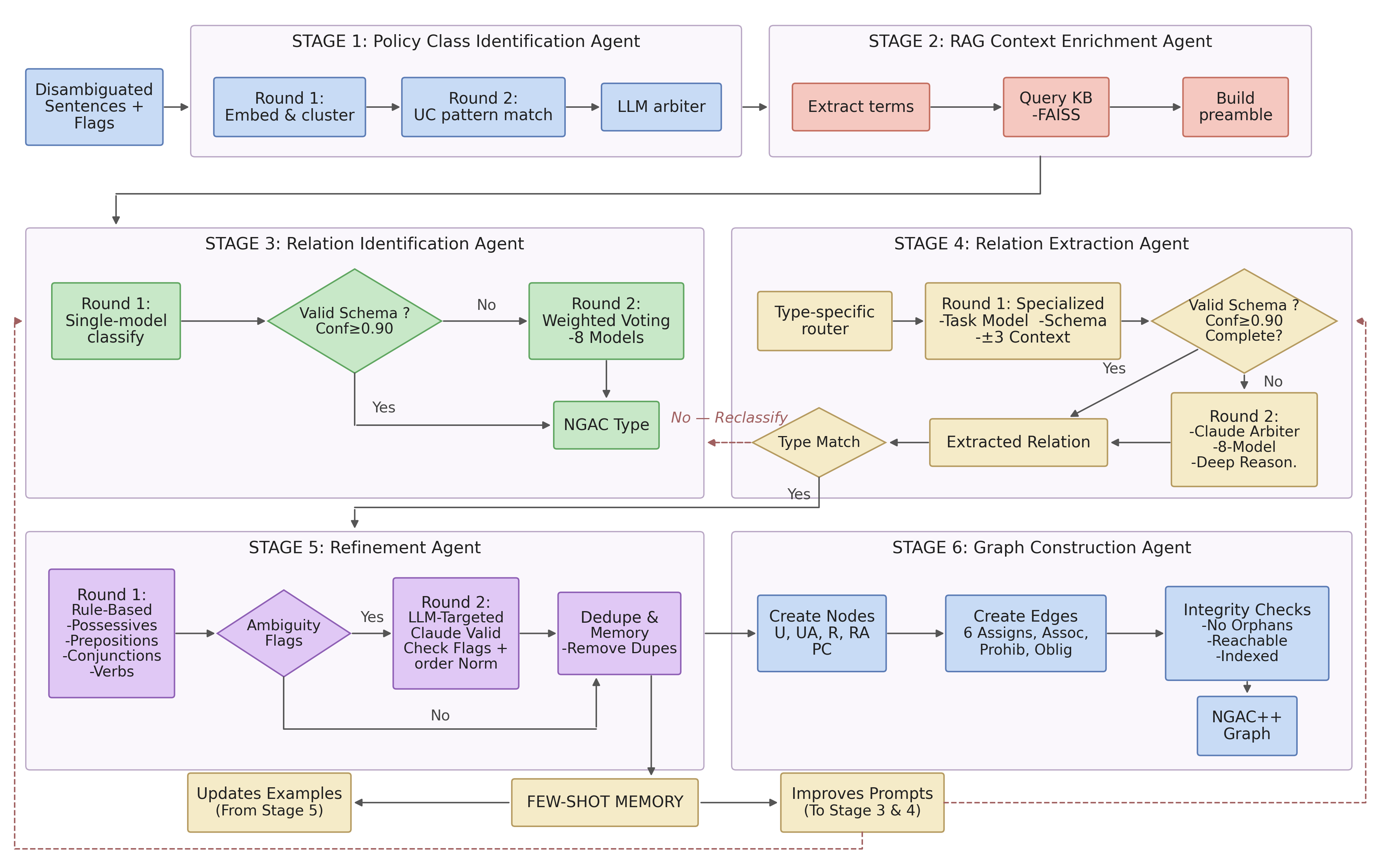}
 \caption{Phase~A six-stage extraction: Stages~1--3 establish context and typing with confidence-gated R1$\rightarrow$R2 escalation; Stages~4--6 produce the graph that Phase~B validates.}
 \label{fig:phaseA}

\end{figure}

Phase~A extracts NGAC tuples from disambiguated sentences through six sequential stages. The first three stages establish context and classification; Stages~4--6 produce the policy graph. Figure~\ref{fig:phaseA} summarizes the data flow.

\subsubsection{Stages 1--2: Policy Class Assignment and RAG Context}
\label{sec:stage1}\label{sec:stage2}

Each sentence is assigned to one of the document's policy classes (PCs). PCs are derived from use-case headers (\eg{} ``UC1: Create and Disable Patients'' yields the PC \texttt{create\_disable\_patients}) by a deterministic preprocessor, then refined: sentences are clustered by sentence-embedding similarity (OpenAI \texttt{text-embedding-3-small}), and an LLM labels each cluster against the candidate PC names. Stage~1 produces the assignment of each ACP sentence to a single PC, which scopes downstream type classification and tuple extraction.

Stage~2 enriches each sentence with retrieval-augmented context drawn from a domain glossary. For iTrust, the glossary defines healthcare-specific terms (\eg{} HCP = ``Health Care Personnel''; MID = ``Medical Identification number''); for CyberChair, conference roles (\eg{} PCC = ``Program Committee Chair''); for IBM, university roles. Glossary terms are embedded with the same embedding model as Stage~1; for each sentence, the top-$k$ most-similar glossary entries are retrieved and prepended to the Stage~3 prompt as context. Stage~2 is the mechanism that allows the same Stage~3 prompt template to generalize across domains without per-dataset prompt engineering.

\subsubsection{Stage 3: Relationship Type Classification (R1 + Confidence-Triggered R2)}
\label{sec:stage3}

Stage~3 classifies each sentence as Association, Prohibition, Obligation, or Assignment, exploiting the per-type model complementarity established in Section~\ref{sec:design-time}. Stage~3 uses a two-round design.

\smallskip\noindent\textbf{R1: single-model classification.} Each sentence is classified by DeepSeek-Chat with a temperature of 0.1 against a structured prompt that includes the disambiguated sentence, the Stage~1 PC assignment, the Stage~2 RAG context, and explicit NGAC type definitions from NIST SP 800-178. The R1 model emits a type label and a confidence score on $\{0.85, 0.95\}$.

\smallskip\noindent\textbf{R2: confidence-gated ensemble.} R1's two-valued confidence makes the 0.9 threshold a binary flag in practice ($\{0.85, 0.95\}$; Section~\ref{sec:threshold-sensitivity}); flagged sentences are escalated to a multi-LLM ensemble that votes across complementary models identified at design time. The Tier-2 trigger rate used in the cost analysis ranges from 3.0\% on iTrust ACRE-n to 20.9\% on iTrust Text2Policy (Section~\ref{sec:cost-effectiveness}); Stage-3 candidate identification is broader once schema-validation failures are included --- 60/234 (25.6\%) on iTrust Text2Policy, 34/54 (63.0\%) on CyberChair, 17/75 (22.7\%) on IBM, and 197/362 (54.4\%) on iTrust ACRE-n --- so the R1$\rightarrow$Final lifts of Table~\ref{tab:stage3} are not bounded by the billable trigger rate (Section~\ref{sec:eval_stage3}). The R2 trigger pattern as a function of dataset size is the subject of an explicit scalability analysis in Section~\ref{sec:r2-scalability}: across the four datasets we find no monotone relationship between trigger rate and corpus size, a finding inconsistent with monotone few-shot scaling at the corpus sizes tested (54--362 sentences).

The threshold sits in a degenerate region, not on a robustness plateau --- R1 emits only two distinct confidence values, so the trigger is effectively binary --- a calibration limitation documented in Section~\ref{sec:threshold-sensitivity}.

\subsubsection{Stage 4: Tuple Extraction}
\label{sec:stage4}

Stage~4 extracts the actual NGAC tuple structure from each classified sentence. The extraction prompt is type-specific: associations and prohibitions extract \texttt{(UA, op, RA)} triples; obligations extract \texttt{(event, response)} pairs where each component is itself a structured tuple; assignments extract \texttt{(child, parent)} pairs and the assignment sub-type. Stage~4 enforces strict NGAC naming conventions (UA/RA prefix, snake\_case, singular form, entity-type tagging).

\subsubsection{Stage 5: Refinement}
\label{sec:stage5}

Stage~5 unifies entity references across sentences. Singular/plural forms (\texttt{patient} vs.\ \texttt{patients}), morphological variants (\texttt{HCP} vs.\ \texttt{health\_care\_provider}), and case differences are normalized through a deterministic ruleset followed by LLM-assisted deduplication. Embedded assignment expansion extracts implicit hierarchical relationships from possessives and prepositions~\citep{slankas2014relation}: \eg{} ``patient's MID'' yields an implicit assignment \texttt{(MID, patient\_record)}.

\subsubsection{Stage 6: NGAC Graph Construction}
\label{sec:stage6}

\begin{figure}[!tbp]
 \centering
 \includegraphics[width=0.65\columnwidth]{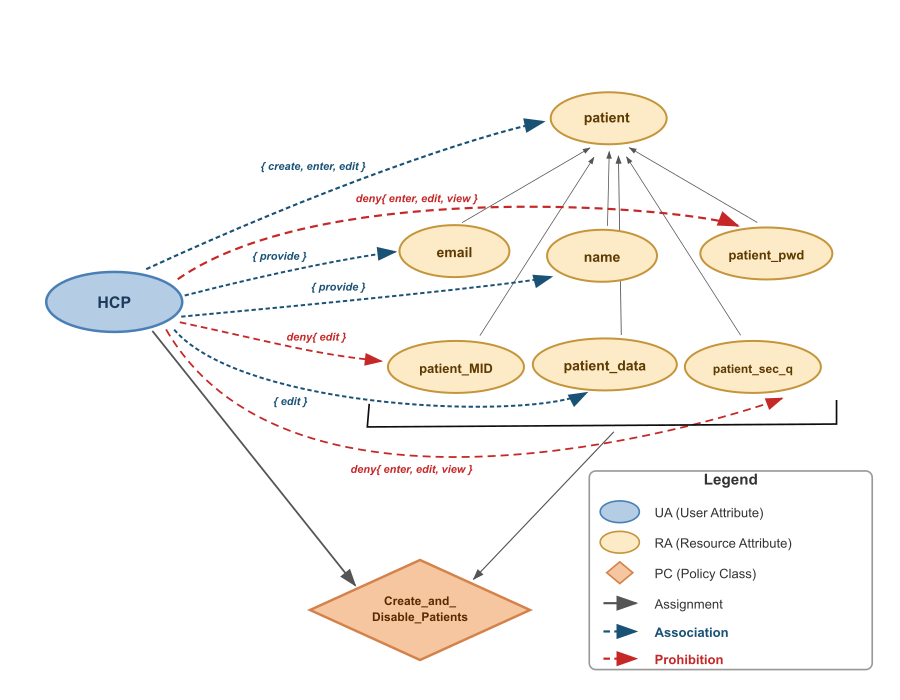}
 \caption{Schematic NGAC graph extracted from iTrust UC1 by Stage~6: assignments (solid black), associations (dashed blue), prohibitions (dashed red), with the contained RA hierarchies and the \texttt{Create\_and\_Disable\_Patients} policy class --- the full released graph for the use case whose simplified subset appears in Figure~\ref{fig:ngac_example}.}
 \label{fig:uc1_graph}

\end{figure}

Stage~6 materializes the refined tuples as a Neo4j property graph. Nodes carry an \texttt{ngac\_type} attribute (one of U, UA, R, RA, PC); edges carry the relationship type (\texttt{ASSIGN}, \texttt{ASSOC}, \texttt{PROHIB}, \texttt{OBLIG}) and the operation. We represent obligations through a paired-edge construction: each obligation contributes a triggering EVENT edge and a corresponding RESPONSE edge, with a shared \texttt{obligation\_id} property linking them. This preserves the full event--response semantics of NIST SP 800-178 while remaining queryable via standard Cypher path expressions. The graph is the canonical artifact handed to Phase~B. Figure~\ref{fig:uc1_graph} shows the schematic graph produced for the iTrust UC1 example; the same graph as rendered by the Neo4j Browser front-end is included in the artifact.

\subsubsection{Running Example: Stage-by-Stage Walkthrough}
\label{sec:walkthrough}

To make Phase~A concrete, Table~\ref{tab:walkthrough} traces the four UC1 sentences (UC1-2, UC1-4, UC1-7, UC1-8) introduced at the start of Section~\ref{sec:approach} through every stage of the pipeline. Compound-sentence decomposition (UC1-4, UC1-8) and implicit-subject inference (UC1-7) happen before extraction, which is what lets Stage~4 produce clean per-conjunct tuples. The same example produces the policy graph rendered in Figure~\ref{fig:ngac_example}.

\begin{table*}[!tbp]
\centering
\caption{Running example: the four UC1 sentences traced through the pipeline.}
\label{tab:walkthrough}
\footnotesize
\setlength{\tabcolsep}{3pt}
\renewcommand{\arraystretch}{1.2}
\resizebox{\textwidth}{!}{%
\begin{tabular}{@{}p{0.5cm}p{2.6cm}p{2.6cm}p{1.7cm}p{1.5cm}p{2.9cm}p{2.4cm}p{2.3cm}@{}}
\toprule
\textbf{ID} & \textbf{Sentence} & \textbf{Phase 0} & \textbf{PC (S1)} & \textbf{Type (S3)} & \textbf{Tuple (S4)} & \textbf{Hierarchies (S5)} & \textbf{Edge (S6)} \\
\midrule
UC1-2 &
``An HCP creates patients.'' &
\emph{(unchanged; ambig.\ score 0.15)} &
\texttt{create\_\allowbreak disable\_\allowbreak patients} &
Association &
\textbf{\texttt{assoc(UA:HCP, create, RA:patient\_record)}} &
$U \to UA$: Bob$\to$HCP &
HCP $-\{c\}-$\newline patient\_record \\
\midrule
UC1-4 &
``Only the name and email are provided by the HCP.'' &
\emph{Score 0.65; conjunction $+$ passive voice $\Rightarrow$} ``The HCP provides the name'' \emph{(and analogously for email)} &
\texttt{create\_\allowbreak disable\_\allowbreak patients} &
Association &
\textbf{\texttt{assoc(UA:HCP, provide, RA:name)}} \;\; \textbf{\texttt{assoc(UA:HCP, provide, RA:email)}} \emph{(one tuple per decomposed branch; verb preserved from the rewrite)} &
$RA \to RA$: name, email contained within patient\_record (extracted from possessive structure) &
HCP $-\{p\}-$\newline name, email \\
\midrule
UC1-7 &
``The patient's MID can not be edited.'' &
\emph{Score 0.40; implicit subject $\Rightarrow$} ``The HCP cannot edit the patient's MID'' &
\texttt{create\_\allowbreak disable\_\allowbreak patients} &
Prohibition &
\textbf{\texttt{prohib(UA:HCP, edit, RA:patient\_mid)}} &
$RA \to RA$: patient\_mid contained within patient\_record (from possessive ``patient's MID'') &
HCP $-\{!e\}-$\newline patient\_mid \\
\midrule
UC1-8 &
``The HCP does not have the ability to enter the patient's security question and password.'' &
\emph{Score 0.55; conjunction $+$ negation $\Rightarrow$} ``The HCP cannot enter the patient's security question'' \emph{(and analogously for password)} &
\texttt{create\_\allowbreak disable\_\allowbreak patients} &
Prohibition (both branches) &
\textbf{\texttt{prohib(UA:HCP, enter, RA:security\_question)}} \;\; \textbf{\texttt{prohib(UA:HCP, enter, RA:password)}} &
$RA \to RA$: security\_question, password contained within patient\_record &
HCP $-\{!ent\}-$\newline security\_question \newline HCP $-\{!ent\}-$\newline password \\
\bottomrule
\end{tabular}%
}
\par\smallskip{\raggedright\footnotesize ``$\rightarrow$'' = Phase-0 decomposition into multiple extraction targets; bolded Stage-4 tuples correspond to the labeled edges of Figure~\ref{fig:ngac_example}; entity names lightly simplified (normalized forms in Figure~\ref{fig:uc1_graph}).\par}
\end{table*}

\subsection{Phase B: Five-Strategy Verification}
\label{sec:phaseB}

Phase~B realizes Contribution~3 and addresses Challenges~C3 (verification gap) and~C4 (verification-rigor gap). Even an accurate extractor produces some errors, and a single mis-classified edge in an NGAC graph can silently introduce a security vulnerability; manual inspection of a 900-edge graph is not a realistic option in deployment. Our design choice is to make verification a first-class pipeline stage, not an afterthought, and to make it \emph{multi-faceted} — the five strategies (coverage, correctness, consistency, obligation completeness, minimality) are deliberately complementary, each catching a different class of extraction error. The accompanying mutation-testing harness (Section~\ref{sec:mutation}) addresses Challenge~C4 by answering a question that no prior NL-to-access-control work answers: ``does the verifier actually catch errors?'' Eight fault-injection operators deliberately corrupt the verified graph, and the fraction Phase~B detects is the empirical measurement of verifier effectiveness. We adapt the mutation methodology of \citet{chen2021mutation} to the NL-extracted-graph setting (scope relation detailed in Section~\ref{sec:related}).

Phase~B verifies the Stage~6 graph against the original sentence corpus through five complementary strategies. A run is reported as PASS only if all five strategies pass their thresholds. One scoping property matters for interpretation: the sentence-grounded tests that S1 executes are themselves LLM-generated from the corpus sentences, so Phase~B's correctness metric is relative to the generated test suite, not to ground truth --- which is precisely why Section~\ref{sec:rigor} evaluates the verifier itself by controlled mutation instead of taking its PASS verdicts at face value.

\smallskip\noindent\textbf{S1: Coverage.} Every ACP sentence in the input must be reflected in at least one tuple in the graph. We compute coverage by checking whether the sentence's identifier appears in the \texttt{sentence\_ids} provenance set of any node or edge. S1 is the check the controlled-mutation evaluation in Section~\ref{sec:mutation} relies on: it provides the sentence-level link between input and output that the mutation harness exploits. (An early version of the pipeline did not propagate \texttt{sentence\_ids} into the operation-merging graph rebuild; we describe the patch we deployed and its empirical effect in Section~\ref{sec:rigor}.)

\smallskip\noindent\textbf{S2: Correctness.} Every node has a well-typed \texttt{ngac\_type}, every edge connects type-compatible endpoints (\eg{} \texttt{ASSOC} edges must run UA $\rightarrow$ RA; \texttt{ASSIGN} edges must respect the six legal assignment sub-types from Definition~5), and every operation is one of the document's declared operations.

\smallskip\noindent\textbf{S3: Consistency.} No two edges $(u, op, r)$ exist as both \texttt{ASSOC} (allow) and \texttt{PROHIB} (deny) for the same triple. NGAC's deny-overrides semantics permits prohibition--association pairs in principle, but unintended pairs indicate extraction error.

\smallskip\noindent\textbf{S4: Obligation completeness.} Every \texttt{OBLIG} edge has a paired EVENT and RESPONSE, both well-typed and connected through their shared \texttt{obligation\_id}.

\smallskip\noindent\textbf{S5: Minimality.} No two edges duplicate the same \texttt{(src, op, tgt, type)}; the operation-merging graph rebuild ensures redundant tuples collapse to a single edge with merged \texttt{sentence\_ids}.

\begin{figure}[!tbp]
 \centering
 \includegraphics[width=\columnwidth]{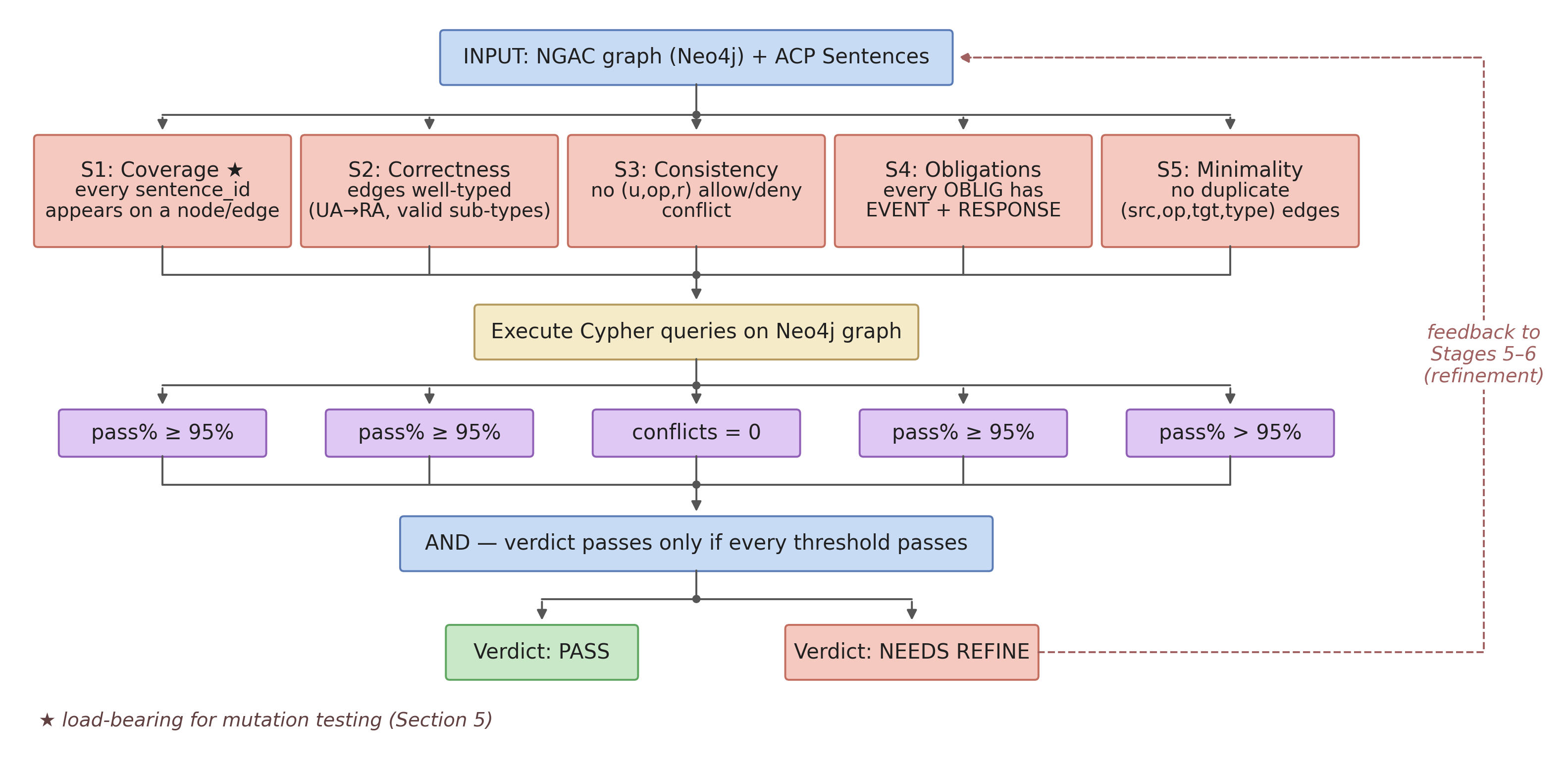}
 \caption{Phase~B five-strategy verification. A run is PASS only if all five strategies pass.}
 \label{fig:phaseB}
\end{figure}

A Phase~B run reports a verdict (PASS / FAIL with detailed reasons), per-strategy pass rates, and the set of sentences and tuples flagged by each failing strategy. We use the verdict in two ways: as the end-of-pipeline correctness check, and as the detection signal in the controlled-mutation evaluation (Section~\ref{sec:mutation}). Figure~\ref{fig:phaseB} summarizes the five-strategy verification flow.

\FloatBarrier
\section{Evaluation}
\label{sec:evaluation}

\noindent This section reports the \emph{extraction-side} results: per-stage accuracy, end-to-end accuracy across the four corpora, reproduced baselines (Abdelgawad, Lawal, BERT-RE), and single-variable ablation. Section~\ref{sec:rigor} reports the \emph{verification-side} results, including mutation testing and its cross-dataset replication.

\subsection{Experimental Design}
\label{sec:exp_design}

RQ1 is answered by per-stage and end-to-end accuracy plus three reproduced baselines (Sections~\ref{sec:eval_stage3}--\ref{sec:baseline}); RQ2 by the ablation and single-model controls (Section~\ref{sec:ablation}); and RQ3 by controlled mutation testing and its cross-dataset replication (Section~\ref{sec:rigor}).

The evaluation uses four datasets (Section~\ref{sec:datasets}) whose ground truth is validated by an independent second-annotator double-coding (Cohen's $\kappa=0.84$ pooled; Section~\ref{sec:datasets}). All metrics are computed against this ground truth using exact tuple match, a strict criterion where partial matches receive no credit. For the baseline comparison (RQ1), original (non-disambiguated) sentences are used to ensure fairness, as the baseline has no equivalent to Phase~0. To quantify sampling uncertainty given the small per-corpus $n$, every reported macro-F1 is accompanied by a 95\% bootstrap confidence interval obtained by resampling sentences with replacement ($B{=}1000$ resamples, percentile method, fixed seed 42); full per-cell intervals appear in the supplementary material.

\subsection{Relation Type Classification}
\label{sec:eval_stage3}

Table~\ref{tab:stage3} presents classification results. The dominant confusion is association $\leftrightarrow$ obligation, where conditional language appears in both types. The low IBM Stage-3 Final macro-F1 and its wide CI reflect small-sample fragility of the rare classes ($n{=}1$ obligation in IBM ground truth) and macro-averaging under class imbalance; most residual IBM misclassifications are corrected by the Stage-4 type-match loop, which lifts IBM type accuracy to 94.7\% (21 sentences reclassified; Section~\ref{sec:eval_stage4}). On iTrust Text2Policy, 60/234 sentences (25.6\%) were routed to Round~2 --- 39 by low confidence, 38 by schema-validation failure, 17 by both --- and R1 was already correct on 52 of the 60, so the ensemble's potential accuracy contribution is concentrated in the 8 R1 errors among them. Per-corpus Tier-2 trigger rates used in the cost model appear in Table~\ref{tab:cost}. 

\begin{table*}[!tbp]
\centering
\caption{Stage~3 type classification and Stage~4 tuple extraction.}
\label{tab:stage3}\label{tab:stage4}
\small
\setlength{\tabcolsep}{4pt}
\begin{tabular}{@{}lrrrrrrrrrr@{}}
\toprule
& \multicolumn{5}{c}{\textbf{Stage 3 (type classification)}} & \multicolumn{5}{c}{\textbf{Stage 4 (tuple extraction)}} \\
\cmidrule(lr){2-6} \cmidrule(lr){7-11}
\textbf{Dataset} & \textbf{R1} & \textbf{Final} & \textbf{P} & \textbf{R} & \textbf{F1} & \textbf{R1} & \textbf{Final} & \textbf{P} & \textbf{R} & \textbf{F1} \\
\midrule
iTrust Text2Policy & 95.3\% & 98.3\% & .983 & .941 & .958 & 80.8\% & 99.6\% & .999 & .999 & .999 \\
iTrust ACRE-n & 74.0\% & 90.6\% & .911 & .950 & .925 & 98.9\% & 99.1\% & .991 & .991 & .991 \\
CyberChair & 64.8\% & 98.1\% & .983 & .989 & .986 & 83.3\% & 98.1\% & .994 & .994 & .994 \\
IBM & 62.7\% & 69.3\% & .650 & .835 & .627 & 84.0\% & 97.3\% & .987 & .980 & .982 \\
\bottomrule
\end{tabular}
\par\smallskip{\raggedright\footnotesize R1 = single model (DeepSeek-Chat); Final = after ensemble R2 --- before type-match feedback for Stage~3, after the full loop for Stage~4. Stage-3 P/R/F1 are macro-averaged over classes; Stage-4 values are per-sentence tuple-set averages; iTrust ACRE-n Stage~4 is scored on $n{=}351$ (11 sentences carry no GT tuple). Per-cell bootstrap CIs in the supplementary material.\par}
\end{table*}

\textbf{Round~1 accuracy and ensemble justification.} The R1 accuracy gap between the design-time benchmark (98\%) and evaluation datasets (62.7\% on IBM, 64.8\% on CyberChair) reflects \emph{systematic per-type bias}. On IBM, most R1 errors are the same misclassification: ASSOCIATION misidentified as OBLIGATION due to conditional language (``If X occurs, the system will Y''). Such systematic biases are what multi-model ensembles and downstream structural evidence resolve: diverse models with different training data do not share the same bias, and an association-format tuple extracted from a sentence classified as obligation is direct structural evidence of the misclassification. On CyberChair the ensemble alone resolves the gap (64.8\%~$\rightarrow$~98.1\%); on IBM the ensemble contributes a smaller share (62.7\%~$\rightarrow$~69.3\%) and the Stage~4 type-match feedback loop corrects a further 21~sentences, reaching 94.7\% type accuracy (Section~\ref{sec:eval_stage4}). On iTrust Text2Policy, where the type distribution matches DeepSeek-Chat's strengths, R1 achieves 95.3\%, leaving the ensemble little to correct. iTrust ACRE-n exhibits a middle pattern: R1 reaches 74.0\% (the larger corpus contains more compound obligations and event-triggered associations that R1 conflates), which the ensemble lifts to 90.6\%. 

\subsection{Tuple Extraction}
\label{sec:eval_stage4}

The Stage-4 block of Table~\ref{tab:stage4} shows extraction results. Table~\ref{tab:stage4_pertype} breaks down iTrust ACRE-n by type: obligation reaches 36/36 (Wilson 95\% CI [90.4\%, 100\%]), and assignment is the weakest class --- its two residual errors are the direction-inversion and containment cases of Table~\ref{tab:qualitative_errors}, which refinement does not repair. Across all four datasets, Stage 4 final accuracy reaches 97.3\%--99.6\% on the corpora's original wording (wording sensitivity: Section~\ref{sec:paraphrase}), demonstrating that the type-match verification mechanism reliably converts Stage~3 classification into well-formed tuples.

\textbf{Stage~4 accuracy exceeding Stage~3 on iTrust Text2Policy and IBM.} On iTrust Text2Policy (99.6\% vs.\ 98.3\%) and IBM (97.3\% vs.\ 69.3\%), Stage~4 final accuracy exceeds Stage~3 final (pre-feedback) accuracy. This is not an error but a consequence of Stage~4's type-match verification: when an extracted tuple's structure is inconsistent with the Stage~3 classification (\eg{} an association-format tuple from an obligation-classified sentence), the mismatch triggers automatic reclassification and re-extraction. On IBM this loop matters: it reclassifies 21~sentences, lifting type accuracy from 69.3\% to 94.7\% before tuple scoring. Consequently, \emph{Stage~3 and Stage~4 are not independent measurements}: in Table~\ref{tab:stage3}, the Stage-3 columns report classification accuracy \emph{before} Stage~4 type-match feedback, while the Stage-4 columns report extraction accuracy \emph{after} the full loop including reclassification. This feedback mechanism is a deliberate design feature that allows downstream extraction evidence to correct upstream classification errors.

\textbf{Controls.} Two potential inflators of the headline are measured and bounded. \emph{In-prompt overlap:} 19/234, 16/54, 1/75, and 4/351 scored sentences (iTrust Text2Policy, CyberChair, IBM, iTrust ACRE-n) appear verbatim in their corpus's own Stage-4 prompt text; all are extracted correctly, and excluding them changes Stage-4 Final accuracy by at most 0.8~pp on any corpus (99.6$\rightarrow$99.5, 98.1$\rightarrow$97.4, 97.3$\rightarrow$97.3, 99.1$\rightarrow$99.1). \emph{Label reliability:} restricting iTrust ACRE-n scoring to the 318 sentences where the independent second annotator agreed with the released label leaves Stage-4 Final accuracy at 99.4\% (316/318), versus 97.0\% on the 33 contested sentences (full treatment in Section~\ref{sec:threats}). Neither test-set leakage nor label ambiguity carries the headline. 

\textbf{Unified end-to-end metric.} Because the two blocks are not independent (the feedback loop above), the Stage-3 and Stage-4 columns of Table~\ref{tab:stage3} cannot be multiplied into an end-to-end number. The end-to-end metric a deployer cares about is sentence-level extraction correctness \emph{after} the full Phase~0--Phase~A extraction pipeline (Phase~B verifies the result rather than correcting it), which equals the Stage~4 ``Final Acc.'' column in Table~\ref{tab:stage4}: iTrust Text2Policy 99.6\%, iTrust ACRE-n 99.1\%, CyberChair 98.1\%, IBM 97.3\%. Stage~3 Final macro-F1 (Table~\ref{tab:stage3}) is reported separately for per-type diagnosis.

\begin{table}[!tbp]
\centering
\caption{iTrust ACRE-n Stage~4: per-type extraction accuracy over the 351 GT-annotated sentences.}
\label{tab:stage4_pertype}
\small
\setlength{\tabcolsep}{10pt}
\begin{tabular}{@{}lrrrr@{}}
\toprule
\textbf{Type} & \textbf{$n$} & \textbf{R1} & \textbf{Final} & \textbf{$\Delta$} \\
\midrule
Association & 236 & 99.2\% & 99.6\% & +0.4\% \\
Assignment & 72 & 97.2\% & 97.2\% & 0 \\
Prohibition & 7 & 100\% & 100\% & 0 \\
Obligation & 36 & 100\% & 100\% & 0 \\
\midrule
\textbf{Total} & \textbf{351} & \textbf{98.9\%} & \textbf{99.1\%} & \textbf{+0.2\%} \\
\bottomrule
\end{tabular}

\end{table}

\subsection{Obligation Extraction}
\label{sec:eval_obligations}

Obligations are the most semantically demanding NGAC type because each tuple requires recognizing an explicit \emph{event} trigger and a \emph{response} action. On every corpus these are \emph{extracted} as structured event(subject, verb, object)$\rightarrow$response tuples (released per corpus in \texttt{ngac\_obligations.csv}). The Stage-6 \emph{paired-edge graph materialization} --- rendering each obligation as an OBLIGATION node with TRIGGERS and REQUIRES scaffolding edges --- is applied to the iTrust Text2Policy graph (12 OBLIGATION nodes, 36 scaffolding edges; Table~\ref{tab:graphs} caption), on which the enforcement replay (Section~\ref{sec:pm-replay}) loads and fires an obligation end-to-end; the other corpora's released Stage-6 graphs retain obligations in tabular form. The paired-edge construction is thus demonstrated and enforced on one corpus, not materialized uniformly across all four. The four datasets contribute 52 obligation sentences in total (Table~\ref{tab:datasets}); on iTrust ACRE-n alone there are 36 obligations, more than the other three datasets combined. Per-corpus obligation counts are small on three datasets; the Wilson intervals carry that uncertainty directly: exact-match obligation accuracy is $9/9$ on iTrust Text2Policy (95\% Wilson CI [70.1\%, 100\%]) and $6/6$ on CyberChair (Wilson CI [61.0\%, 100\%]), while iTrust ACRE-n, with $n{=}36$ sentences at $36/36$ (Wilson CI [90.4\%, 100\%]), provides the tightest evidence that the obligation pipeline is correct on real corpora. 

\textbf{Sentence-to-tuple mapping.} The 9~obligation sentences in iTrust Text2Policy yield 12~obligation tuples because some sentences specify multiple event triggers within a single policy statement. UC201 (``Alert by Email'') contains one sentence listing four distinct triggering events (password change, lab procedure status change, comparison report request, comparison report generation), each producing a separate obligation tuple with the same response. This is a many-to-one expansion at the tuple level, not over-extraction: each tuple corresponds to a distinct event--response pair explicitly described in the source text. The same many-to-one expansion holds for iTrust ACRE-n at larger scale.

\subsection{NGAC Graph Statistics}
\label{sec:eval_graphs}

Upstream of the graphs, Stage~1 derives 36/44/14/8 policy classes per corpus (iTrust Text2Policy / iTrust ACRE-n / CyberChair / IBM) and Stage~2's retrieval knowledge bases hold 61/64/60/15 domain terms. 

Table~\ref{tab:graphs} summarizes the generated policy graphs. The iTrust ACRE-n graph (411 nodes, 969 edges) is larger than the IBM and CyberChair graphs combined, providing direct evidence that the pipeline scales structurally to richer real-world policies. The 44 policy classes in iTrust ACRE-n (vs.\ 8--36 in the other datasets) also stress-test the policy-class assignment stage (Stage~1).

\smallskip\noindent\textbf{Assignment sub-type coverage (measured).} The six NIST SP 800-178 assignment sub-types are exercised unevenly by real policy documents. Counting assignment edges per sub-type in the released Stage-6 edge CSVs across iTrust Text2Policy / CyberChair / IBM / iTrust ACRE-n: RA$\rightarrow$PC 280/71/55/303, UA$\rightarrow$PC 66/25/24/110, R$\rightarrow$RA 47/11/1/40, RA$\rightarrow$RA 35/22/1/130, UA$\rightarrow$UA 2/0/0/4, and U$\rightarrow$UA \textbf{0/0/0/0}. On the first three corpora these sums reconcile exactly with Table~\ref{tab:graphs} (Edges $-$ Assoc.\ $-$ Deny $=$ 430/129/81), and every assignment edge carries one of the five occurring legal sub-types. On iTrust ACRE-n the released graph's 660 assignment edges exceed this census (587) by 73; the topology audit classifies every residual delta as an entity$\rightarrow$PC anchoring edge added by the final graph rebuild, with all non-PC sub-type counts matching exactly, so no assignment edge falls outside the legal sub-types. Five of the six sub-types occur in the corpora; \emph{individual-user} assignment (U$\rightarrow$UA, binding a named person to an attribute) does not appear, because these documents specify attribute- and role-level rules, not assignments of named individuals. The pipeline's representation supports U$\rightarrow$UA --- the running-example figure includes illustrative user nodes (Alice, Bob) that are additions, not extractions --- but we report the claim as \emph{five of six sub-types exercised}, not all six. 

\begin{table}[!tbp]
\centering
\caption{Generated NGAC policy graph statistics, counted from the released Stage~6 graph CSVs.}
\label{tab:graphs}
\small
\setlength{\tabcolsep}{10pt}
\begin{tabular}{@{}lrrrrr@{}}
\toprule
\textbf{Dataset} & \textbf{Nodes} & \textbf{Edges} & \textbf{PC} & \textbf{Assoc.} & \textbf{Deny} \\
\midrule
iTrust Text2Policy & 321 & 662 & 36 & 224 & 8 \\
iTrust ACRE-n & 411 & 969 & 44 & 299 & 10 \\
CyberChair & 90 & 167 & 14 & 38 & 0 \\
IBM & 57 & 135 & 8 & 49 & 5 \\
\bottomrule
\end{tabular}
\par\smallskip{\raggedright\footnotesize Edges = ASSIGNMENT $+$ ASSOCIATION $+$ PROHIBITION after operation merging (one edge per source--target--type triple); iTrust Text2Policy's 36 obligation-scaffolding REQUIRES/TRIGGERS edges are excluded (Section~\ref{sec:limitations}).\par}
\end{table}

\subsection{Phase B: Verification Results}
\label{sec:eval_phaseb}

Table~\ref{tab:phaseb} presents Phase~B results from the development-time run, scored against the LLM-generated S1 suite (Section~\ref{sec:phaseB}); the deterministic re-runs on the released graphs appear in Table~\ref{tab:rigor_summary} (Section~\ref{sec:replication}). Three of the four datasets achieve PASS with $\geq$99.4\% test pass rates; IBM reports NEEDS REFINE on a single genuine consistency conflict, the temporal-scoping catch documented in Section~\ref{sec:ibm-case} and preserved as a diagnostic case study, not fixed. 

\begin{table*}[!tbp]
\centering
\caption{Phase~B verification results from the \emph{initial development-time run}.}
\label{tab:phaseb}\label{tab:phaseb_diagnosis}
\small
\setlength{\tabcolsep}{4pt}
\begin{tabular}{@{}lrrrrrrrrrrr@{}}
\toprule
& \multicolumn{6}{c}{\textbf{Verification run}} & \multicolumn{5}{c}{\textbf{Failure diagnosis vs.\ GT}} \\
\cmidrule(lr){2-7} \cmidrule(lr){8-12}
\textbf{Dataset} & \textbf{Tests} & \textbf{Pass\%} & \textbf{Compl.} & \textbf{Corr.} & \textbf{Con.} & \textbf{Min.} & \textbf{Fail} & \textbf{TP} & \textbf{FP} & \textbf{Missed} & \textbf{Recall} \\
\midrule
iTrust Text2Policy & 1430 & 99.4\% & 100\% & 98.6\% & 0 & 97.7\% & 9 & 1 & 8 & 0 & 100\% \\
iTrust ACRE-n & 1987 & 99.6\% & 98.9\% & 99.2\% & 0 & 97.4\% & 8 & 1 & 0 & 0 & 100\% \\
CyberChair & 287 & 99.7\% & 100\% & 99.1\% & 0 & 95.8\% & 1 & 1 & 0 & 0 & 100\% \\
IBM & 418 & 97.4\% & 100\% & 95.4\% & 1 & 99.2\% & 11 & 2 & 9 & 0 & 100\% \\
\bottomrule
\end{tabular}
\par\smallskip{\raggedright\footnotesize Compl./Corr./Con./Min.\ = completeness / correctness / consistency conflicts / minimality; TP = genuine error; FP = Cypher formatting noise; Missed = GT errors not caught. Recall = TP / genuine GT errors flagged in this run (1, 1, 1, 2 per dataset); the E3-class parent-abstraction error on iTrust Text2Policy (Section~\ref{sec:taxonomy-ops}) is structurally undetectable and outside this denominator. TPs include graph-level defects not scored in the Stage-4 tuple census of Section~\ref{sec:eval_errors} (iTrust Text2Policy's TP is the S5 minimality catch among the 16 redundant transitive assignments cleaned before release, Section~\ref{sec:replication}); iTrust ACRE-n's containment-naming and direction-inversion residuals produce well-typed, conflict-free edges outside the suite's observables and are scored only in the Stage-4 census; the 7 iTrust ACRE-n failures not shown as TP or FP are boundary mismatches reclassified as verification-correct (Section~\ref{sec:cypher-fp}).\par}
\end{table*}

\textbf{Verification diagnostic analysis.} To substantiate Phase~B's detection capability, we cross-validated all test failures against ground truth on all four datasets (Table~\ref{tab:phaseb_diagnosis}). Each failure is categorized as: \emph{true positive} (genuine extraction error confirmed by GT), \emph{false positive} (Cypher formatting noise: entity-name casing, underscore conventions), or \emph{missed error} (GT error not flagged by Phase~B).

The verifier's diagnostic value depends on both recall and precision. \emph{Recall:} no genuine GT errors were missed on any dataset ($5/5$). \emph{Precision:} the initial-run FP rate is high on iTrust Text2Policy and IBM, the two corpora processed before the entity-normalization patches ($17$ of $20$ flags were Cypher formatting noise), which the three pipeline patches of Section~\ref{sec:cypher-fp} eliminate ($0\%$ FP on the post-patch runs); the S1-provenance pre-patch pair (iTrust Text2Policy and CyberChair) was preserved in its original state to enable the $2{\times}2$ patch-validation comparison of Section~\ref{sec:mutation}, and deployment readiness should be judged on the post-patch precision, since the patches ship in the released pipeline.

\textbf{Phase~B as a diagnostic instrument: case study from iTrust ACRE-n.} Phase~B's first run on iTrust ACRE-n achieved 98.9\% completeness, 99.2\% correctness, and 97.4\% minimality, but flagged a single same-operation conflict on \texttt{(UA:hcp, edit, RA:patient\_mid)}, blocking a clean PASS verdict. Rather than dismiss this as a borderline case, we treated it as a diagnostic signal and traced its provenance. The conflict arose from two tuples drawn from different sentences: from sentence ID 75 (``Patient MID should be the number assigned when the patient is added to the system and cannot be edited'') came the correct prohibition \texttt{deny(UA:hcp, edit, RA:patient\_mid)}; from sentence ID 150 (``HCP must enter the MID of a patient and then enter or edit demographic information$\ldots$'') came the spurious association \texttt{assoc(UA:hcp, edit, RA:patient\_mid)}. The second extraction is wrong: in the coordinated phrase ``enter or edit demographic information'', the verb ``edit'' attaches to ``demographic information'', not to ``MID''. The Stage~4 model fell into a classic coordinated verb-phrase attachment ambiguity, propagating ``edit'' leftward to the nearer noun. Both edges entered the graph, and Phase~B's S3 consistency check correctly identified the same-operation grant/deny pair as a violation. We removed the spurious tuple and added the correct one (\texttt{assoc(UA:hcp, edit, RA:demographic\_information)}), then re-ran Phase~B; the conflict resolved to zero, yielding a clean PASS verdict on all four properties. This case demonstrates Phase~B's value beyond rubber-stamp validation: the structural consistency check (S3) caught a single-tuple extraction error that would otherwise have silently introduced a contradictory policy into a 969-edge graph. Manual inspection of 362 ACP sentences would not reliably surface such localized cross-sentence inconsistencies; automated graph-level verification does so by construction.

\subsection{Error Analysis}
\label{sec:eval_errors}

\begin{figure}[!tbp]
 \centering
 \includegraphics[width=0.78\columnwidth]{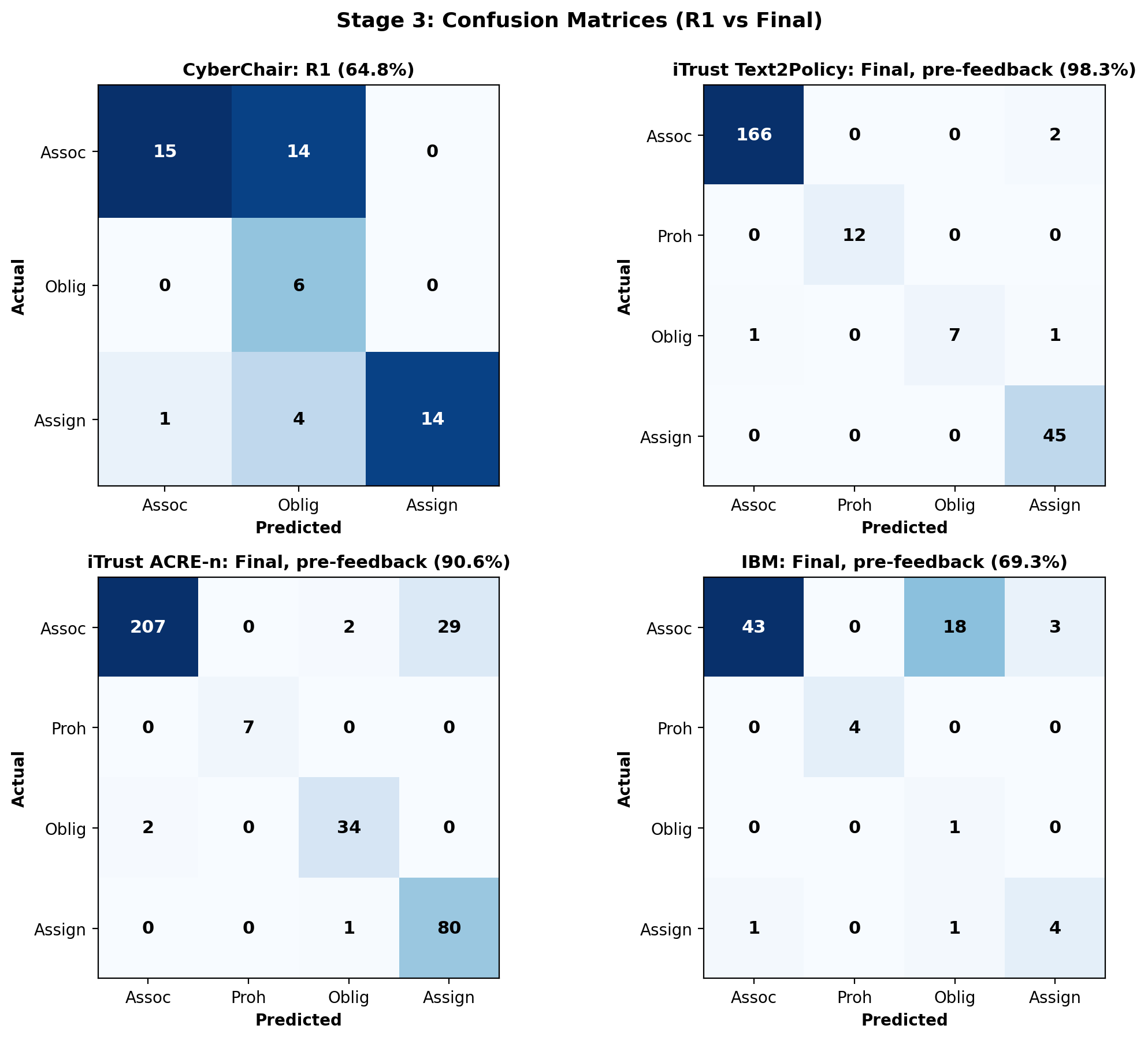}
 \caption{Stage~3 confusion matrices (rows = ground truth, columns = predicted; Final matrices use the pre-feedback convention of Table~\ref{tab:stage3}).}
 \label{fig:confusion}

\end{figure}

Across the four datasets, the pipeline produces 7~total errors out of the 714 scored sentences: 1 on iTrust Text2Policy, 3 residual on iTrust ACRE-n, 1 on CyberChair, and 2 on IBM. These cluster into four diagnostic classes: (1) UA/RA prefix confusion, (2) coordinated-VP attachment ambiguity, (3) parent-entity abstraction, and (4) CRUD-style type-boundary confusion. Each is the subject of a Section~\ref{sec:replication} case study or a documented residual. The three residual iTrust~ACRE-n errors are one VP-attachment error (the patient\_mid case, Section~\ref{sec:itrust-case}), one containment-naming error (\texttt{RA:mid} extracted with the wrong parent), and one assignment-direction error (a UA-to-UA assignment in the wrong direction); iTrust Text2Policy's single error is the parent-abstraction case (\texttt{LHCP\_message\_inbox}, Table~\ref{tab:qualitative_errors}), which is E3-class and structurally undetectable by design (Section~\ref{sec:taxonomy-ops}). Of the seven, the patient\_mid conflict was caught by Phase~B's S3 without human pre-screening (Section~\ref{sec:eval_phaseb}). Table~\ref{tab:qualitative_errors} provides one canonical example per pattern, and Figure~\ref{fig:confusion} shows the Stage~3 confusion matrices in which CyberChair R1's systematic Association$\rightarrow$Obligation bias is corrected by the ensemble. Pattern~2 (coordinated-VP attachment) has no row: it manifests as the cross-sentence \texttt{patient\_mid} grant/deny conflict of Section~\ref{sec:eval_phaseb}, not as a single-tuple mismatch. The two rows marked \emph{residual} are the containment-naming and assignment-direction errors on iTrust ACRE-n.

\begin{table*}[!tbp]
\centering
\caption{Qualitative error catalogue: one canonical example per diagnostic pattern.}
\label{tab:qualitative_errors}
\footnotesize
\setlength{\tabcolsep}{4pt}
\renewcommand{\arraystretch}{1.25}
\resizebox{\textwidth}{!}{%
\begin{tabular}{@{}p{0.085\textwidth}p{0.25\textwidth}p{0.235\textwidth}p{0.25\textwidth}p{0.10\textwidth}@{}}
\toprule
\textbf{Dataset} & \textbf{Sentence (truncated)} & \textbf{Ground truth tuple} & \textbf{\system{} prediction} & \textbf{Pattern} \\
\midrule
CyberChair & ``\ldots{} the PCC may ask non-experts to assign \ldots{} reviewers \ldots{}'' & \texttt{assoc(UA:PCC,}\newline\texttt{ask, RA:reviewer)} & \texttt{assoc(UA:PCC,}\newline\texttt{ask, UA:reviewer)} & 1: UA/RA prefix \\
iTrust ACRE-n & ``\ldots{} the personal representative may view test results on behalf of the patient.'' & \texttt{assoc(UA:representative,}\newline\texttt{view,}\newline\texttt{RA:test\_result)} & \texttt{assoc(UA:representative,}\newline\texttt{view,}\newline\texttt{UA:test\_result)} & 1: UA/RA prefix \\
iTrust ACRE-n & ``The HCP cannot edit the patient's MID.'' & \texttt{prohib(UA:HCP,}\newline\texttt{edit,}\newline\texttt{RA:patient\_mid)} & \texttt{prohib(UA:HCP, edit,}\newline\texttt{RA:mid)}\newline (incorrect parent) & residual: containment naming \\
iTrust Text2Policy & ``\ldots{} the LHCP receives messages in the message inbox.'' & \texttt{assoc(UA:LHCP,}\newline\texttt{read, R:system)} & \texttt{assoc(UA:LHCP, read,}\newline\texttt{RA:LHCP\_message\_inbox)} & 3: parent abstraction \\
iTrust ACRE-n & ``LHCPs are HCPs with limited privileges.'' & \texttt{assign(UA:LHCP}\newline\texttt{$\to$UA:HCP)} & \texttt{assign(UA:HCP}\newline\texttt{$\to$UA:LHCP)}\newline (direction inverted) & residual \\
IBM & ``This use case allows the registrar to register students for courses.'' & \texttt{assoc(UA:registrar,}\newline\texttt{register,}\newline\texttt{RA:course)} & \texttt{assign(R:course\_reg\_system}\newline\texttt{$\to$RA:course)} & 4: type boundary \\
\bottomrule
\end{tabular}%
}

\end{table*}

\subsection{Baseline Comparison}
\label{sec:baseline}

We reproduced the approach of \citet{abdelgawad2023synthesizing} using their three classification algorithms with spaCy's \texttt{en\_core\_web\_lg} model on all four datasets. To ensure a fair comparison, we used original (non-disambiguated) sentences, as the baseline has no equivalent to Phase~0 semantic disambiguation. Both the Abdelgawad reproduction and the Lawal Code4Policy reproduction received identical input (raw sentences plus the same use-case header context that \system{} consumes), unlimited retry budget on transient failures, and the same ground-truth annotations for scoring; the comparison measures algorithmic capability under matched-input conditions, not budget or pre-processing differences. Table~\ref{tab:baseline} presents the comparison. The reproduction is faithful within the baseline's design scope: restricted to association-typed sentences, the class Abdelgawad \etal's rules target, it reaches 91.1\% on iTrust Text2Policy ($n{=}168$), 75.6\% on iTrust ACRE-n ($n{=}238$, matching their reported $\sim$75\%), 44.8\% on CyberChair ($n{=}29$), and 48.4\% on IBM ($n{=}64$) --- the lower CyberChair and IBM in-scope figures reflect those corpora's non-CRUD verb vocabulary. The macro-F1 gap on the full corpora arises predominantly from the type classes outside the rule-set's scope (assignment recall 0.049) and therefore quantifies scope, not reproduction infidelity. 

\begin{table}[!tbp]
\centering
\caption{Unified baseline comparison on the four real corpora.}
\label{tab:baseline}\label{tab:baseline-lawal}\label{tab:baseline-bert}
\small
\setlength{\tabcolsep}{4pt}
\begin{tabular}{@{}lcccccccc@{}}
\toprule
& \multicolumn{3}{c}{\textbf{Abdelgawad et al.}} & \multicolumn{2}{c}{\textbf{Lawal Code4Policy}} & \textbf{BERT-RE} & \multicolumn{2}{c}{\textbf{\system{}}} \\
\cmidrule(lr){2-4} \cmidrule(lr){5-6} \cmidrule(lr){7-7} \cmidrule(lr){8-9}
\textbf{Dataset} & P & R & F1 & Type-F1 & Tuple-F1 & macro-F1 (OOD) & Type-F1 & Tuple-F1 \\
\midrule
iTrust Text2Policy & .624 & .431 & .477 & 0.447 & 0.211 & 0.608 & 0.958 & 0.999 \\
iTrust ACRE-n & .372 & .454 & .367 & 0.335 & 0.206 & --- (train) & 0.925 & 0.991 \\
CyberChair & .310 & .246 & .192 & 0.177 & 0.047 & 0.422 & 0.986 & 0.994 \\
IBM & .530 & .457 & .492 & 0.485 & 0.234 & 0.291 & 0.627$^{\ddagger}$ & 0.982 \\
\bottomrule
\end{tabular}
\par\smallskip{\raggedright\footnotesize Abd.\ P/R/F1 = macro precision/recall/F1; Type-F1 = classification macro-F1; Tuple-F1 = per-sentence tuple-set F1; BERT-RE is trained on iTrust ACRE-n and evaluated out-of-domain only. $^{\ddagger}$Pre-feedback Stage-3 macro-F1; the type-match loop lifts IBM to 0.867 (Section~\ref{sec:eval_stage4}). Per-cell bootstrap CIs in the supplementary material.\par}
\end{table}

The baseline's accuracy is severely limited by its rule-based dependency parsing: compound sentences, passive voice, and pronoun chains all produce extraction failures. Its highest F1 (0.492 on IBM) is achieved on the corpus with the simplest sentence structure; on CyberChair, where assignment hierarchies are deepest, F1 collapses to 0.192. Across the four datasets, \system{}'s per-sentence tuple F1 is 0.982--0.999 against the reproduction's macro F1 of 0.192--0.492; the two columns are different aggregations of identical ground truth (see the table footnote), so we report them side by side, not as a subtraction. The reproduction is weakest exactly where the rule-based parser fails most often: CyberChair (0.192) and iTrust ACRE-n (0.367).

\subsubsection{Comparison to Lawal et al.\ Code4Policy}
\label{sec:baseline-lawal}

The Abdelgawad baseline is rule-based; a fairer LLM-vs-LLM comparison is the more recent Code4Policy approach of \citet{lawal2024translating}, which prompts an LLM to instantiate a Python \texttt{@dataclass} schema representing access-control tuples. Lawal \etal\ report 99\% F1 on LLM-synthesized policy documents but did not evaluate on real-world corpora. We reproduce Code4Policy on all four of our datasets using the exact \texttt{@dataclass} schema and few-shot prompts from their Listings~1 and~2, instantiated on DeepSeek-Chat at temperature~0 --- the same base model and endpoint as \system{}'s own R1 extraction --- so the comparison isolates the extraction \emph{method} rather than confounding it with model capability. 

\textbf{Different scopes by design, and the structural cap that follows.} Lawal \etal{} scoped Code4Policy to CRUD-style associations and prohibitions: the published \texttt{@dataclass} schema defines only CRUD operation codes ($\{r, w, x, c, d\}$ and their negations) and cannot express ASSIGNMENT or OBLIGATION tuples, whereas \system{} extracts all four NGAC types. Because assignment- and obligation-typed sentences account for 9\%--46\% of ACPs per corpus (28\% pooled), the type-F1 of \emph{any} Code4Policy reproduction is upper-bounded near $0.5$ regardless of which LLM instantiates the schema; the numbers below quantify the cost of CRUD-only scope on full-NGAC corpora, not a deficiency in Code4Policy within its intended scope.

\textbf{Metrics.} We report \emph{type-F1} (macro-F1 across the four NGAC type classes, with assignment and obligation contributions structurally zero under the cap) and \emph{tuple-F1} (exact-match F1 against the full ground truth after symmetric verb canonicalization through Code4Policy's CRUD codes plus object-side plural lemmatization). Table~\ref{tab:baseline-lawal} reports the head-to-head results.

\textbf{Interpretation.} \system{}'s Tuple-F1 and Abdelgawad's macro-F1 aggregate identical ground truth differently (per-sentence vs.\ macro), so Table~\ref{tab:baseline} reports them side by side, not as a subtraction. The cleanest comparison is same-scope, same-metric: restricted to Code4Policy's own design scope (association- and prohibition-typed sentences only, $n{=}29$--$245$ per corpus), its per-sentence tuple-F1 is 0.310/0.267/0.058/0.233 on iTrust Text2Policy/iTrust ACRE-n/CyberChair/IBM against \system{}'s 1.000/0.996/0.989/0.995 on the identical subsets and metric (0.058--0.310 vs.\ 0.989--1.000) --- a gap scope cannot explain. In full, under the structural cap just described, Code4Policy's type-F1 lands at $0.18$--$0.49$ and its tuple-F1 at $0.05$--$0.23$ (CyberChair's 0.047 is the mismatch at its sharpest: 25 of 54 sentences carry schema-inexpressible ground truth, and most remaining association sentences use conference-domain verbs outside the CRUD vocabulary). We also checked whether the gap is a scoring artifact from entity-naming conventions: re-scoring the baseline under entity-side normalization (possessive-\emph{s} stripping, plural lemmatization, underscore collapse) raises its tuple-F1 by only 2--5~pp (\eg{} CyberChair 0.05$\rightarrow$0.07, IBM 0.23$\rightarrow$0.29), closing a small fraction of the $\sim$0.7 gap and confirming it reflects extraction failure, not naming penalties. 
Most of the gap to Lawal's published $0.99$ (on synthesized, schema-co-designed data) is thus explained partly by scope and partly by this in-scope gap, and \system{}'s advantage comes from covering all four types plus the disambiguation, retrieval, and ensemble machinery of Phases~0 and~A. We did not re-run Code4Policy on Lawal's three synthesized corpora (BoA, NHS, UTSA): their ground truth is by construction expressible in the CRUD-only schema and free of natural-language ambiguity, so the run would reproduce the published $0.99$ essentially by construction while answering none of the generalization questions the real corpora answer; moreover, those synthesized corpora are not publicly archived, so a confirmatory reproduction on them requires the original authors' data.

\subsubsection{Comparison to a Fine-Tuned BERT-RE Baseline}
\label{sec:baseline-bert}

The Abdelgawad and Lawal baselines are, respectively, rule-based and prompting-based. The remaining obvious alternative is the IE/RE literature's default, a fine-tuned relation-extraction model, which we add as a third baseline: \texttt{bert-base-uncased} (110M parameters) fine-tuned for NGAC type classification. We choose iTrust ACRE-n as the training corpus because it is the largest of our four real corpora ($n{=}362$), is the only corpus with non-trivial obligation samples ($n{=}36$), and is the canonical stress-test dataset in our evaluation. Training is 5 epochs, batch~16, learning rate $2{\times}10^{-5}$ (the standard full-fine-tuning rate for \texttt{bert-base}),\footnote{A learning-rate sweep under the identical protocol confirms the configuration is not under-tuned and the conclusion is rate-robust; cross-domain macro-F1 never exceeds 0.52 on CyberChair or IBM at any rate (the supplementary material). 
} max length~128, with weighted cross-entropy by inverse class frequency to address rare-class imbalance; seed 42, fully deterministic. End-to-end training plus evaluation completes in 6.3~s on a single RTX~5090. 
We report \emph{out-of-domain} evaluation only: in-domain evaluation would measure memorization, and the meaningful comparison with \system{}'s zero-shot pipeline is cross-corpus transfer. Table~\ref{tab:baseline-bert} reports the per-corpus result.

\textbf{Interpretation.} Two observations matter for the comparison:
(i)~\emph{Out-of-domain transfer is poor.} Macro-F1 lands at $0.422$ on CyberChair (a different conference-domain corpus), $0.291$ on IBM (different educational domain), and $0.608$ on iTrust Text2Policy (the same healthcare domain as ACRE-n but a different curation). One out-of-domain hop costs the fine-tuned model 23--55 percentage points of macro-F1 from its in-domain 0.842 (in-domain measures memorization and is reported only as the reference point for the drop; per-class values ship with the artifact). This is the structural reason a fine-tuned RE pipeline cannot serve as a turn-key NGAC extraction stage on heterogeneous corpora.
(ii)~\emph{Per-class collapse on rare types.} BERT-RE scores $0.000$ F1 on prohibition for all three out-of-domain datasets and on obligation for two of three: the small absolute counts of those classes ($n{=}1$--$12$ in the OOD corpora, $n{=}7$--$36$ in training) leave too little signal for the model to learn the rare classes, even with weighted cross-entropy.

\system{}, by contrast, is zero-shot at inference time and reaches type macro-F1 of $0.925$--$0.986$ on three corpora and $0.867$ on IBM after the type-match loop (Table~\ref{tab:baseline}), including the three OOD corpora used for BERT-RE evaluation. The baseline therefore strengthens the case for the LLM-driven pipeline: BERT converged on its $n{=}362$ training sentences yet does not transfer across the corpus boundary.

\subsection{Ablation Study}
\label{sec:ablation}

Table~\ref{tab:ablation} presents the ablation results. \textbf{Design.} The four single-model conditions --- R1 with full context, $-$RAG, $-$Phase~0, and both removed --- were all run under one pinned DeepSeek-Chat snapshot (temperature 0.1), using each corpus's verbatim Stage-4 prompt assembly (system prompt with domain glossary and NGAC rules, per-type few-shots, $\pm$3-sentence context, recorded Stage-3 type routing). Because all four arms share one snapshot, the within-table deltas are free of vendor drift. The $-$RAG condition is a surgical excision, not a prompt rewrite: it omits only the per-sentence retrieved domain-context block while retaining the policy-class line, the surrounding-sentence context, and the disambiguated sentence. Ensemble evidence is reported in a separate table (Table~\ref{tab:ablation_ensemble}), grouped by snapshot, because cross-snapshot comparisons compound component effects with vendor drift. The ablation covers iTrust Text2Policy, CyberChair, and IBM. Instrumentation health per cell: zero API failures; at most 9 of 234 unparseable responses, so the magnitudes below are not artifacts of broken outputs. Statistical testing uses exact McNemar tests on paired per-sentence exact-match outcomes. 

\begin{table*}[!tbp]
\centering
\caption{Ablation study: matched single-snapshot arms.}
\label{tab:ablation}
\small
\begin{tabular}{@{}lccc@{}}
\toprule
\textbf{Configuration} & \textbf{iT.~T2P} & \textbf{Cyber.} & \textbf{IBM} \\
\midrule
R1 single model (re-run) & 58.5 / .638 & 87.0 / .888 & 58.7 / .644 \\
$-$ RAG (Stage 2) & 58.5 / .636 & 85.2 / .867 & 60.0 / .651 \\
$-$ Phase 0 & 51.3 / .568 & 87.0 / .892 & 61.3 / .657 \\
$-$ RAG $-$ Phase 0 & 52.6 / .593 & 87.0 / .904 & 58.7 / .628 \\
\midrule
$\Delta$ RAG (vs.\ R1) & 0.0~pp ($p{=}1.0$) & $-$1.9~pp ($p{=}1.0$) & $+$1.3~pp ($p{=}1.0$) \\
$\Delta$ Phase 0 (vs.\ R1) & $\mathbf{-7.3}$~pp ($p{=}0.002$) & 0.0~pp ($p{=}1.0$) & $+$2.7~pp ($p{=}0.75$) \\
$\Delta$ both (vs.\ R1) & $-$6.0~pp ($p{=}0.009$) & 0.0~pp ($p{=}1.0$) & 0.0~pp ($p{=}1.0$) \\
\bottomrule
\end{tabular}
\par\smallskip{\raggedright\footnotesize Cells: exact-match \% / mean tuple-F1 ($n$ = 234/54/75). $\Delta$ rows: within-run change vs.\ the same-run R1 arm with exact McNemar $p$-values; all arms share one pinned snapshot (drift-free deltas). $\Delta$ values are computed on per-sentence counts before rounding and can differ from differences of the rounded percentages by 0.1~pp.\par}
\end{table*}

\begin{table}[!tbp]
\centering
\caption{Ensemble evidence, grouped by model snapshot.}
\label{tab:ablation_ensemble}
\small
\begin{tabular}{@{}lccc@{}}
\toprule
\textbf{Configuration (exact-match \%)} & \textbf{iT.~T2P} & \textbf{Cyber.} & \textbf{IBM} \\
\midrule
\multicolumn{4}{@{}l}{\emph{Archived snapshot (production record)}} \\
\quad R1 single model & 80.8 & 83.3 & 84.0 \\
\quad Full pipeline & 99.6 & 98.1 & 97.3 \\
\midrule
\multicolumn{4}{@{}l}{\emph{Current snapshot (re-run toggled arms)}} \\
\quad R1 DeepSeek-Chat & 58.5 & 87.0 & 58.7 \\
\quad R1 DeepSeek-Reasoner & 59.0 & 81.5 & 57.3 \\
\bottomrule
\end{tabular}

\end{table}

\textbf{What the measured ablation shows.} (i)~\emph{The ensemble-plus-refinement layer dominates, and its contribution is stable across snapshots.} On the archived same-snapshot record, single-model R1 Stage-4 accuracy is 80.8\%/83.3\%/84.0\% against the full pipeline's 99.6\%/98.1\%/97.3\% (Table~\ref{tab:ablation_ensemble}; per-corpus record in Table~\ref{tab:stage4}), a drift-free +13.3 to +18.8~pp; this is a pipeline-stage comparison recorded in production, not a toggled ablation arm, and the current-snapshot comparison below provides the toggled-arm corroboration. Under the current snapshot, the archived full pipeline beats the re-run R1 arm by +11.1 to +41.0~pp, with \emph{every} discordant sentence pair favouring the full pipeline (McNemar $p \leq 0.031$ on all three corpora); this second comparison compounds ensemble benefit with vendor drift and is corroborative only. Per-class, IBM's rare classes collapse entirely without the ensemble: the R1 arm scores 0/6 on assignments, 1/4 on prohibitions, and 0/1 on obligations, so the full pipeline's rare-class correctness on IBM is supplied by the ensemble-plus-refinement layer, not by the single model. (ii)~\emph{Phase~0 has a significant single-model effect where the single model is weakest.} On iTrust Text2Policy, removing Phase~0 costs 7.3~pp ($p{=}0.002$; the degradation concentrates in associations, 61.9\%$\rightarrow$52.4\%, and assignments); on CyberChair and IBM the single-model effect is null. (iii)~\emph{RAG shows no measurable single-model effect on any corpus under the current snapshot} ($-$1.9 to $+$1.3~pp, all $p{=}1.0$). Full per-condition per-class accuracy and F1 are in the released artifact (\texttt{ablation\_summary.json}).

\textbf{Component contributions are snapshot-dependent.} The archived ablation runs against the earlier vendor snapshot measured much larger single-model contributions: removing RAG cost 20.4~pp on CyberChair (83.3\%$\rightarrow$63.0\%) and 62.4~pp on iTrust Text2Policy (80.8\%$\rightarrow$18.4\%), with Phase-0 effects of similar magnitude, and the missing glossary produced the systematic Association$\rightarrow$Obligation R1 bias visible in Figure~\ref{fig:confusion}. 
The current snapshot reproduces neither effect: it reaches 87.0\% on CyberChair with no pipeline support at all. The parsimonious reading is that the newer model has internalized much of the disambiguation and domain grounding that the earlier snapshot obtained from Phase~0 and RAG. Two consequences follow. First, the layered design functions as \emph{insurance whose value depends on the underlying model}: largest where the model is weakest (every archived-snapshot effect, and the Phase-0 effect on iTrust Text2Policy today), null where the model has caught up. Second, single-model component contributions measured on one vendor snapshot do not transfer to another --- an instance of the vendor-drift threat (Section~\ref{sec:threats}). The one contribution stable across both snapshots is the ensemble-plus-refinement layer.

\textbf{Single-model controls.} To rule out the objection that the ensemble's value is an artifact of a weak base model, we ran the identical R1 configuration on DeepSeek-Reasoner and on two other vendors' frontier models --- GPT-5 (\texttt{gpt-5-2025-08-07}) and Claude-Sonnet-5 (\texttt{claude-sonnet-5}) on stratified 40-sentence subsets per corpus, with GPT-4o-mini (\texttt{gpt-4o-mini-2024-07-18}) additionally on the full corpora; exact identifiers are pinned here and in the artifact. No single model significantly outperforms DeepSeek-Chat as a single extractor on any corpus (paired exact McNemar $p \geq 0.09$ everywhere), and against the full pipeline the gap is one-directional: no single-model arm ever solved a sentence the pipeline missed (discordant pairs 3:0 to 122:0; $p < 0.001$ on every corpus wide enough to power the test). Absolute single-arm levels carry the same cross-snapshot caveat as the ablation; the within-subset paired direction does not. The headline accuracy therefore comes from the layered ensemble-plus-refinement machinery, not raw single-model strength. 

\subsection{Wording-Robustness Probes}
\label{sec:paraphrase}

Extraction accuracy on a fixed benchmark says nothing about sensitivity to how the same policy is worded --- a robustness dimension the NL-to-policy literature does not normally measure. We measure it with controlled paraphrase probes at three levels --- the single-model extraction path (three corpora), the complete architecture on a substitute same-family roster, and the complete architecture on the production roster --- plus a semantic re-scoring of all arms that separates true extraction failure from surface-form scoring artifacts. Table~\ref{tab:paraphrase} collects the results; the probes also serve the training-data-contamination analysis in Section~\ref{sec:limitations}.

\noindent\textbf{Method.} Per corpus, 40 ground-truth-bearing ACP sentences (stratified by NGAC type, seed~42) were paraphrased by DeepSeek-Chat at temperature~0.7 under instructions to preserve every actor, action, resource, polarity, and event--response structure; DeepSeek-Reasoner screened each paraphrase for semantic preservation, discarding 4/2/4 of 40 on iTrust Text2Policy, CyberChair, and IBM. Both arms of each pair were extracted by the corpus's verbatim machinery and scored by exact tuple-set match with paired bootstrap CIs ($B{=}1000$). For the semantic score, every strict mismatch passed deterministic normalization, batched LLM adjudication (DeepSeek-Chat, $T{=}0$), and a hand review of every verdict under a conservative rubric: synonym operations and same-referent renames count as correct; type or polarity changes, actor or resource referent changes (including dropped owner prefixes), missing or extra rules, and empty predictions never do. All paraphrases, per-sentence records, adjudication verdicts, and review overrides ship with the artifact. 

\begin{table*}[!tbp]
\centering
\caption{Wording-robustness probes: paired original-vs-paraphrase deltas (percentage points, 95\% bootstrap CIs).}
\label{tab:paraphrase}
\footnotesize
\setlength{\tabcolsep}{5pt}
\begin{tabular}{@{}llrrccc@{}}
\toprule
\textbf{Probe} & \textbf{Corpus} & \textbf{Orig.} & \textbf{Para.} & \textbf{$\Delta$ strict} & \textbf{$\Delta$ op-wildcarded} & \textbf{$\Delta$ semantic} \\
\midrule
Single path (R1) & iTrust Text2Policy & 58.3 & 8.3 & $-50.0$ $[-66.7, -36.1]$ & $-36.1$ $[-52.8, -22.2]$ & $-19.4$ $[-33.3, -8.3]$ \\
 & CyberChair & 89.5 & 47.4 & $-42.1$ $[-60.5, -23.7]$ & $-26.3$ $[-42.1, -10.5]$ & $-18.4$ $[-34.2, -5.3]$ \\
 & IBM & 58.3 & 22.2 & $-36.1$ $[-52.8, -22.2]$ & $-25.0$ $[-38.9, -11.1]$ & $-13.9$ $[-30.6, +2.8]$ \\
\midrule
Full pipeline, substitute roster & CyberChair & 60.5 & 26.3 & $-34.2$ $[-50.0, -21.1]$ & --- & --- \\
Full pipeline, production roster & CyberChair & 47.4 & 21.1 & $-26.3$ $[-42.1, -10.5]$ & --- & $-7.9$ $[-21.1, +5.3]$ \\
\bottomrule
\end{tabular}
\par\smallskip{\raggedright\footnotesize Orig./Para.\ = strict exact-match accuracy per arm. Semantic-arm accuracies: single-path originals 80.6/92.1/66.7\% vs.\ paraphrases 61.1/73.7/52.8\%; production pipeline 50.0$\rightarrow$42.1\%. Substitute roster: DeepSeek-Chat base, DeepSeek-Reasoner escalation/arbitration; production roster: Claude Phase-0 validation, the reduced three-model R2 ensemble of Section~\ref{sec:cost-effectiveness}, GPT-4o-mini second extraction, Claude arbitration.\par}
\end{table*}

\noindent\textbf{Reading the table.} The paired deltas, not the absolute levels, are the comparable quantity: the arms run against current vendor snapshots (CyberChair reproduces its archived R1 within 3.7~pp; iTrust Text2Policy drifts by roughly $-22$~pp; Section~\ref{sec:ablation}) and, for the single path, without the R2 ensemble or Stage-5 refinement --- but both arms of each pair share the drift equally. Two mechanical artifacts inflate the strict deltas: Stage~4's verb-preservation rule copies the paraphrase's verb (\eg{} \emph{submit} where the original said \emph{send}), which string-mismatches ground truth anchored to the original verb (the op-wildcarded column removes this), and 14 of 38 scored CyberChair sentences appear verbatim in that corpus's own few-shot prompt (1 and 0 on the others) --- excluding them leaves the op-wildcarded deltas essentially unchanged, so the residual drop is distinguishable from zero on all three corpora. Three findings follow. The single-model path is materially sensitive to meaning-preserving rewording, and roughly half of its strict-match collapse is scoring artifact, not extraction failure. The layered architecture recovers only $\sim$8~pp of the loss under the substitute roster. Under the production roster the semantic delta of $-7.9$~pp carries a wide interval at $n{=}38$ --- consistent with anything from no drop to a $\sim$20-pp drop --- so the probe bounds the full pipeline's paraphrase sensitivity rather than establishing wording-invariance. The residual drop cannot be attributed to pretraining memorization specifically: few-shot examples, prompt conventions, and ground truth are all anchored to the original phrasing, and paraphrasing shifts the input distribution, each of which depresses the paraphrase arm independently of memorization. Reported headline accuracies should accordingly be read as accuracy on the corpora's original wording.

\subsection{Efficiency Analysis}
\label{sec:efficiency}

Table~\ref{tab:efficiency} reports per-dataset wall-clock time and measured API spend. Wall-clock scales linearly with corpus size (0.55--0.64 minutes per sentence across a 6.7$\times$ size range), dominated by serial LLM calls; the only cross-sentence processing dependency is the run-time few-shot memory, and the trigger-rate analysis (Section~\ref{sec:r2-scalability}) finds no monotone benefit from its accumulation at the corpus sizes tested, so throughput should parallelize across sentences with little accuracy cost --- relevant for enterprise-scale documents, where a 10{,}000-sentence corpus would take $\sim$4 days serially; validating parallel equivalence at that scale is untested and noted as future work. For enterprise-scale documents, per-sentence processing is independent, amortized API cost extrapolates linearly (\$11--\$47 per 1{,}000 sentences; Section~\ref{sec:cost-effectiveness}), and the trigger-rate analysis found no size-driven growth in escalation over the 6.7$\times$ size range tested --- though validation on corpora beyond $\sim$400 sentences remains future work. Two cost figures appear in this paper and measure different things: (i)~the \emph{measured gross spend} of Table~\ref{tab:efficiency} is the total API expenditure over each corpus's full processing history, including schema-validation re-prompts, transient-failure retries, and partial re-runs; its per-sentence quotient (\$0.09--\$0.12) does not track the Tier-2 trigger rate for exactly this reason and should not be read as a marginal cost. (ii)~The \emph{amortized marginal cost} of Table~\ref{tab:cost} models a clean single pass as \$0.005 + trigger rate $\times$ \$0.20 per sentence (\$0.011--\$0.047 across the four corpora) and is the deployment-relevant figure. The 2.6--11$\times$ gap between the two is development overhead, not hidden marginal cost.

\begin{table}[!tbp]
\centering
\caption{End-to-end pipeline efficiency across the four datasets.}
\label{tab:efficiency}
\small
\setlength{\tabcolsep}{10pt}
\begin{tabular}{@{}lrrrrr@{}}
\toprule
\textbf{Dataset} & \textbf{ACPs} & \textbf{R2 \%} & \textbf{Wall-clock} & \textbf{Gross spend (devel.-inclusive)} \\
\midrule
iTrust Text2Policy & 234 & 20.9\% & $\sim$2.5 h & $\sim$\$28 \\
iTrust ACRE-n & 362 & 3.0\% & $\sim$3.5 h & $\sim$\$45 \\
CyberChair & 54 & 14.8\% & $\sim$30 m & $\sim$\$5 \\
IBM & 75 & 16.0\% & $\sim$45 m & $\sim$\$8 \\
\bottomrule
\end{tabular}

\end{table}

\FloatBarrier
\section{Mutation-Based Verification Rigor}
\label{sec:rigor}

\noindent\textbf{Experimental provenance.} Results in this section come from four measurement states named consistently throughout: the \emph{development-time run} (pre-cleanup working graphs, Table~\ref{tab:phaseb}), the \emph{released graphs} (published Stage-6 outputs, on which every deterministic re-run executes), \emph{pre- vs.\ post-patch} verifier states (S1 provenance patch toggled on identical graphs), and \emph{archived vs.\ current model snapshots}; table captions name the state they report.

We adapt the NGAC mutation methodology of \citet{chen2021mutation} to a narrower question: whether Phase~B's verifier reliably detects corruptions of an \emph{NL-extracted} graph whose upstream extractor (Phases 0 and A) is itself fallible (Section~\ref{sec:related} gives the full positioning). This section establishes the empirical evidence through four complementary experiments (mutation testing, R2 trigger-rate scalability analysis, Cypher false-positive rate analysis, and test-suite coverage analysis), then closes with a sensitivity analysis of the pipeline's two key thresholds and an end-to-end cost-effectiveness analysis.

\subsection{Mutation Testing of Phase B}
\label{sec:mutation}

Following the mutation analysis methodology of \citet{chen2021mutation}, we defined eight mutation operators, each targeting a specific Phase~B verification strategy. Table~\ref{tab:mutation_ops} formalizes the operators.

\begin{table}[!tbp]
\centering
\caption{Eight mutation operators with formal definitions.}
\label{tab:mutation_ops}
\small
\begin{tabular}{@{}p{0.05\textwidth}p{0.70\textwidth}p{0.18\textwidth}@{}}
\toprule
\textbf{Op} & \textbf{Operation} & \textbf{Targets} \\
\midrule
M1 & Remove a uniformly-random ASSOCIATION edge & S1 Coverage \\
M2 & Add a spurious ASSOCIATION between a random unrelated UA--RA pair (empty \texttt{sentence\_ids}) & S1 Provenance \\
M3 & Replace one operation on a random edge with a different operation from the operation pool & S2 Correctness \\
M4 & Add a same-operation grant where a deny exists, creating a conflict & S3 Consistency \\
M5 & Reverse the direction of a random ASSOCIATION edge (UA$\rightarrow$RA $\Rightarrow$ RA$\rightarrow$UA) & S2 Correctness \\
M6 & Re-label a random ASSOCIATION as a PROHIBITION (preserving direction and ops) & S3 / S2 \\
M7 & Add an exact-duplicate ASSOCIATION edge (parallel edge with identical attributes) & S5 Minimality \\
M8 & Replace an obligation tuple with a truncated version having an empty response field & S4 Obligations \\
\bottomrule
\end{tabular}
\par\smallskip{\raggedright\footnotesize A mutant is detected if any Phase~B metric (correctness, conflicts, test failures, minimality) moves in the direction predicted by the operator's target strategy.\par}
\end{table}

\textbf{Experimental protocol.} For each dataset, we generated 30 mutants per operator (with random seed 42 fixed for reproducibility) from the clean Phase~B-verified graph. A mutant is counted as detected if any Phase~B metric deviates in the direction predicted by the operator's target strategy. M4 trials are skipped when the random sample picks a (UA, RA) pair that already has a coexisting same-operation association in the baseline; these are vacuous trials, not failures. The wall-clock cost of the experiment is approximately 18~seconds per dataset on a single CPU; no LLM API calls are required, making the experiment fully reproducible at zero marginal cost. \emph{Seed sensitivity:} re-running the complete harness under five seeds (42--46) varies the aggregate post-patch structural detection between 98.7\% and 99.9\% and the taxonomy-suite aggregate between 72.6\% and 75.2\%; all reported values use seed~42, and per-seed records ship with the artifact. 

\textbf{The S1 edge-provenance patch.} The first run of the experiment surfaced a coverage gap in Phase~B's original S2 strategy: M2 mutants (spurious added edges) were undetectable because S2 tests are sentence-derived, and a random spurious edge has no sentence backing. We patched Phase~B by extending S1 with an \emph{edge-provenance} requirement: every ASSOCIATION and PROHIBITION edge must carry a non-empty \texttt{sentence\_ids} list. This requires the Stage~6 graph rebuild to propagate \texttt{sentence\_ids} from each contributing tuple to the merged edge, a small but essential change. The patched check raises M2 detection from 0\% to 100\% across all post-patch datasets while introducing zero baseline failures (every naturally-extracted edge carries provenance). The patch could equivalently be framed as fixing an implementation gap in the original Phase~B pipeline that we shipped without detecting; mutation testing is the harness that surfaced the gap. In both framings, a controlled-injection harness is what exposes coverage gaps a verifier cannot detect on its own output.

\textbf{Per-dataset results.} Table~\ref{tab:mutation_results} reports detection rates for each operator and dataset, in both the historical state and the uniform post-patch re-run; the controlled 2$\times$2 M2-detection split of the historical state (100\% post-patch vs.\ 0\% pre-patch with no exceptions) independently corroborates the patch effect: the four datasets span three domains and two graph sizes, so the only systematic difference between the two groups is the patch itself. The deterministic before/after toggle below is the primary evidence for the patch; the natural experiment serves as independent corroboration.

\begin{table*}[!tbp]
\centering
\caption{Mutation detection rates by operator and dataset (post-patch verifier).}
\label{tab:mutation_results}
\footnotesize
\setlength{\tabcolsep}{4pt}
\begin{tabular}{@{}lcccc@{}}
\toprule
\textbf{Op} & \textbf{iT.~ACRE-n} & \textbf{IBM} & \textbf{CyberChair} & \textbf{iT.~T2P} \\
\midrule
M1 & 100\% & 100\% & 96.7\%$^{\ddagger}$ & 100\% \\
M2 (original run) & \textbf{100\%} & \textbf{100\%} & \textbf{0\%} & \textbf{0\%} \\
M2 (post-patch) & \textbf{100\%} & \textbf{100\%} & \textbf{100\%} & \textbf{100\%} \\
M3 & 100\% & 100\% & 100\% & 100\% \\
M4 & 100\% & 100\% & NA & 100\% \\
M5 & 100\% & 100\% & 93.3\%$^{\ddagger}$ & 100\% \\
M6 & 100\% & 100\% & 100\% & 100\% \\
M7 & 100\% & 100\% & 100\% & 100\% \\
M8 & 100\% & 100\% & 100\% & 100\% \\
\midrule
\textbf{Aggregate (original run)} & \textbf{237/237} & \textbf{231/231} & \textbf{177/210 (84.3\%)} & \textbf{210/240 (87.5\%)} \\
\textbf{Aggregate (post-patch)} & \textbf{237/237} & \textbf{231/231} & \textbf{207/210 (98.6\%)} & \textbf{240/240} \\
\bottomrule
\end{tabular}
\par\smallskip{\raggedright\footnotesize $n{=}30$ seeded mutants per cell (seed 42, deterministic). $^{\ddagger}$Reproducible fault masking on the single baseline-flagged CyberChair edge ($n{=}300$ probe; Section~\ref{sec:mutation}). M2 rows: historical split vs.\ uniform post-patch re-run (same graphs, same seed); M4 NA: CyberChair has no PROHIBITION edges. Aggregates count completed trials; fractions without percentages are 100\%.\par}
\end{table*}

\textbf{Aggregate.} Across all four datasets, 918 mutants complete (vacuous and inapplicable M4 trials excluded). In the historical mixed pre-/post-patch state, Phase~B detects 855 of them (93.1\%), with every miss confined to M2 on the two pre-patch datasets (60 mutants) plus 3 CyberChair mutants that landed on a single baseline-flagged defective edge (fault masking; footnote to Table~\ref{tab:mutation_results}). Under the uniform post-patch re-run described next, detection is 915/918 (99.7\%). 

\smallskip\noindent\textbf{Fault masking and the uniform re-run.} The three residual misses expose a property of mutation testing itself: \emph{detection is conditional on a clean baseline}. Corruptions stacked on an edge already implicated in a baseline-flagged extraction error move no metric and pass silently; a targeted 2{,}400-trial probe detected every one of its 2{,}323 healthy-edge mutations, and the masking vanishes once the flagged defect is corrected (the same NEEDS-REFINE workflow applied to the patient\_mid conflict). Because the verifier, the Stage-6 rebuild, and the mutation harness are deterministic and CPU-only, we re-ran the verifier side uniformly on all four datasets, rebuilding the two pre-patch corpora from archived per-sentence tuples with \texttt{sentence\_ids} propagated; toggling the S1 provenance patch off and on over identical graphs flips M2 detection 0\%$\rightarrow$100\%, isolating the patch effect from vendor drift. 

\subsection{Taxonomy-Derived Fault Operators}
\label{sec:taxonomy-ops}

The eight operators above were each designed to target a specific Phase~B strategy, so the operator set and the verifier are co-designed (Section~\ref{sec:limitations}): 100\% detection of faults built to exercise the verifier's own checks is necessary but not sufficient evidence of effectiveness against arbitrary extraction faults. To break the circularity, we defined four additional operators sampled from an independent source, the empirical Stage~4 error taxonomy of Section~\ref{sec:eval_errors}: \textbf{E1} retypes an association target with the wrong UA/RA prefix (the PCC$\rightarrow$reviewer class); \textbf{E2} reattaches an operation to a sibling resource in the same RA hierarchy (the coordinated-VP patient\_mid class); \textbf{E3} replaces an association target with its hierarchy parent (the parent-entity abstraction class); \textbf{E4} re-labels an association as a containment ASSIGNMENT (the CRUD-style type-boundary class). Unlike M1--M8, every E-mutant edge \emph{keeps} its \texttt{sentence\_ids}, because a real extractor fault carries provenance, so detection cannot lean on the S1 provenance check and is scored only on movement in correctness, conflicts, or minimality. The experiment is deterministic and CPU-only (30 mutants per operator per dataset, seed 42, run against the post-patch verifier); Table~\ref{tab:taxonomy_ops} reports the results. E2 is inapplicable on CyberChair and IBM, whose graphs contain no sibling-bearing RA hierarchies (analogous to M4-NA on CyberChair). 

\begin{table}[!tbp]
\centering
\caption{Taxonomy-derived fault operators: detection by dataset.}
\label{tab:taxonomy_ops}
\small
\setlength{\tabcolsep}{8pt}
\begin{tabular}{@{}lccccc@{}}
\toprule
\textbf{Op (error class)} & \textbf{iT.~ACRE-n} & \textbf{IBM} & \textbf{CyberChair} & \textbf{iT.~T2P} & \textbf{Overall} \\
\midrule
E1 UA/RA prefix confusion & 30/30 & 30/30 & 29/30 & 30/30 & 119/120 (99.2\%) \\
E2 coordinated-VP attachment & 28/30 & NA & NA & 30/30 & 58/60 (96.7\%) \\
E3 parent-entity abstraction & 8/30 & 0/30 & 0/30 & 3/30 & \textbf{11/120 (9.2\%)} \\
E4 CRUD type-boundary & 30/30 & 30/30 & 27/30 & 30/30 & 117/120 (97.5\%) \\
\bottomrule
\end{tabular}
\par\smallskip{\raggedright\footnotesize E-mutants keep sentence provenance, so detection is scored on correctness/conflict/minimality movement only. NA: no sibling-bearing RA hierarchies in the corpus graph.\par}
\end{table}

Three of the four independently-derived fault classes are detected at 96.7--99.2\%, materially strengthening the effectiveness claim beyond the co-designed operator set. The E3 result is a measured detection boundary: \textbf{a parent-entity abstraction fault passes verification silently in roughly nine cases out of ten} (across five seeds, E3 detection spans 9.2--17.5\% and never exceeds 18\%, so the boundary is seed-robust), so a real extractor error of this class would reach deployment undetected. Three points characterize the boundary. \emph{Mechanism:} an edge moved from \texttt{patient\_mid} to its parent \texttt{patient\_record} remains a well-typed, provenance-carrying, conflict-free edge, and Phase~B's hierarchy-aware access resolution, the same NGAC-correct inheritance semantics defended in Section~\ref{sec:cypher-fp}, resolves queries about the child through the parent, legitimizing the fault by construction; the 11 detected E3 cases are those where the operation merge happened to move a correctness or conflict metric. \emph{Consistency:} Section~\ref{sec:eval_errors} already classified parent-entity abstraction errors as ``both defensible''; E3 turns that qualitative gesture into a quantitative limit. \emph{Implication:} graph-structural verification has a measured boundary at semantically-plausible abstraction-level faults; deployments requiring protection against this class need sentence-grounded checks rather than additional structural rules. A deterministic pilot of exactly this kind --- flag any association whose target attribute has a hierarchy descendant named in the edge's own source sentence --- recovers 86.7/100/100/93.3\% of the E3 mutants on iTrust Text2Policy/CyberChair/IBM/iTrust ACRE-n (114/120 overall) while flagging only 2.3--5.3\% of healthy association edges for expert review. It is a review-routing heuristic, not an automated repair --- some healthy flags are legitimately parent-targeted edges whose sentences mention sub-attributes --- but it shows the E3 boundary is addressable with sentence grounding at low review cost; the detector and per-mutant records ship with the artifact. 

The E3 boundary is \emph{structural}, inherent to hierarchy-aware graph verification; the single E1 miss and all three E4 misses are \emph{contingent} fault masking on one baseline-flagged edge and vanish once it is corrected, while E2's two misses are genuine semantic absorption (per-mutant traces ship with the artifact). 

\subsection{Security-Consequence Weighting}
\label{sec:severity}

Counting detected faults treats all corruptions alike, but their security consequences are asymmetric. A fault that removes or restricts legitimate access \emph{fails closed}: an availability defect, disruptive but safe. A fault that grants access that should not exist, or broadens a privilege beyond its intended scope, \emph{fails open}: a latent policy violation. Table~\ref{tab:severity} classifies the twelve operators by consequence and re-reports detection accordingly (post-patch verifier, as-extracted baselines). The four observed extraction-error classes (Section~\ref{sec:eval_errors}) map the same way: UA/RA prefix confusion produces dangling references (fail-closed), the IBM temporal-scoping conflict over-restricts (fail-closed), and parent-entity abstraction broadens privilege (fail-open).

\begin{table}[!tbp]
\centering
\caption{Detection re-reported by security consequence (post-patch verifier).}
\label{tab:severity}
\small
\begin{tabular}{@{}p{0.21\textwidth}p{0.5\textwidth}r@{}}
\toprule
\textbf{Consequence} & \textbf{Operators (fault effect)} & \textbf{Detection} \\
\midrule
Fail-open, structural & M2 spurious grant; M3 operation substitution; M4 conflict grant; M8 obligation-response loss & 438/438 (100\%) \\
Fail-open, semantic & E2 operation reattachment; E3 parent-entity privilege broadening; E4 on the non-canonical flagged edge (inheritance-broadening) & 69/183 (37.7\%) \\
Fail-closed & M1 edge deletion; M5 direction reversal; M6 grant$\rightarrow$deny flip; E1 UA/RA retype (dangling); E4 grant$\rightarrow$containment flip (canonical edges) & 593/597 (99.3\%) \\
Redundancy hygiene & M7 duplicate edge & 120/120 (100\%) \\
\bottomrule
\end{tabular}
\par\smallskip{\raggedright\footnotesize M3 is classified by its granting component. E4 is split empirically: the injected containment edge adds no reachable grant on any canonical edge, but becomes a legal UA$\rightarrow$UA assignment on the one baseline-flagged edge, whose inheritance broadens privilege.\par}
\end{table}

Two facts emerge. First, every \emph{structural} fail-open fault is detected: 438/438 across spurious grants (M2), operation substitutions (M3), conflict grants (M4), and obligation-response losses (M8). Second, the residual fail-open exposure concentrates entirely in the \emph{semantically plausible} classes: E2's two undetected reattachments, E4's three undetected mutants of the non-canonical flagged edge (whose grant$\rightarrow$containment flip yields a legal UA$\rightarrow$UA assignment that lets \texttt{PCC} inherit \texttt{reviewer}'s grants, verified by exhaustive reachability analysis over all 610 association edges), and, dominantly, E3's privilege broadening (109 of 120 undetected). The verifier's one measured weak point is precisely its highest-severity class: a parent-entity abstraction fault silently broadens privilege in roughly nine of ten cases. Concretely, an association that should target \texttt{patient\_mid} but is abstracted to its parent \texttt{patient\_record} grants the operation on \emph{every} sub-attribute the parent contains --- security question, password, MID --- through ordinary NGAC inheritance, and the resulting edge is well-typed, provenance-carrying, and conflict-free, so it passes every structural check. The exposure does not require an attacker: an ordinary extraction error of this class reaches deployment as a silent privilege broadening, and a source document whose wording invites abstraction (possessives spanning several sub-attributes) makes the error more likely. The available mitigations today are procedural --- flagging hierarchy-parent targets whose source sentence names a specific sub-attribute for expert review --- while the automated fix, sentence-grounded semantic checks, is the concrete future-work item this boundary motivates. Fail-closed faults, whose consequence is availability, not violation, are detected at 99.3\% (593/597). 

\subsection{R2 Trigger-Rate Scalability Analysis}
\label{sec:r2-scalability}

Across the four corpora the R2 trigger rate shows no monotone relationship with corpus size (per-quartile slopes disagree in direction, all $p \geq 0.057$), inconsistent with few-shot scaling at the sizes tested ($n{=}54$--$362$); it tracks sentence-type difficulty instead. Full quartile analysis is in the supplementary material.

\subsection{Cypher False-Positive Rate Analysis}
\label{sec:cypher-fp}

The original Phase~B implementation reported an 85\% Cypher false-positive rate (17/20) on iTrust Text2Policy and IBM --- the two corpora processed before the entity-normalization patches, a different pair from the S1-provenance pre-patch pair of Section~\ref{sec:mutation} --- dominated by entity-name normalization noise (Table~\ref{tab:phaseb_diagnosis}). Three pipeline patches --- strict operation matching in S2, the S1 edge-provenance check (Section~\ref{sec:mutation}), and an operation-preserving Stage-6 rebuild --- eliminate this class: the post-patch Cypher FP rate is 0\% on all corpora (on iTrust ACRE-n, of 8 residual Phase~B failures 1 is the patient\_mid true positive and 7 are verification-correct boundary mismatches from UA-ancestor inheritance). The deterministic re-runs on the released graphs are FP-free (Table~\ref{tab:rigor_summary}); patch details and the per-failure breakdown are in the supplementary material.

\subsection{Test-Suite Coverage Analysis}
\label{sec:coverage}

Phase~B's test suite covers 93.9\%--100\% of NGAC structural elements per class across the four datasets (operation-triple coverage 98.0\%--100\%; aggregate 97.7\%--100\%). The one iTrust ACRE-n gap is a single uncovered grant triple targeting the flagged non-canonical UA$\rightarrow$UA edge Phase~B has already flagged for review. The per-element breakdown is in the supplementary material. 

\subsection{Cost-Effectiveness}
\label{sec:cost-effectiveness}

Deployment cost matters for a pipeline built on commercial LLMs; three design choices bound it. First, we separate the pipeline into a mandatory \textbf{Tier~1} (Phase~0 ambiguity scoring + Phase~A R1 extraction + Phase~B verification, costing $\sim$\$0.005/sentence using DeepSeek) and an optional \textbf{Tier~2} (R2 multi-model ensemble + Claude Opus (\texttt{claude-opus-4-20250514}) arbitration, costing $\sim$\$0.20/sentence on triggered sentences only). Across the four paper datasets, the Tier~2 R2 trigger rate ranges from 3.0\% (iTrust ACRE-n) to 20.9\% (iTrust Text2Policy), yielding amortized effective costs of \$0.011--\$0.047 per sentence end-to-end, translating to \$11--\$47 per 1{,}000-sentence policy document. Second, we conducted a reduced-ensemble experiment in which the full five-model R2 ensemble was replaced with a smaller three-model configuration (DeepSeek-Reasoner, GPT-4o-mini, and Claude Sonnet (\texttt{claude-sonnet-4-5-20250514}), retaining the best per-type performers from the design-time benchmark in Section~\ref{sec:preliminaries}). Third, we identify open-source fine-tuning as a concrete future-work path to reduce cost further.

\begin{table*}[!tbp]
\centering
\caption{End-to-end pipeline cost across four datasets.}
\label{tab:cost}
\small
\setlength{\tabcolsep}{10pt}
\begin{tabular}{@{}lcccc@{}}
\toprule
\textbf{Dataset} & \textbf{Tier 2 trig.} & \textbf{Full-ens. cost} & \textbf{Reduced cost} & \textbf{$\Delta$ F1} \\
\midrule
iTrust ACRE-n & 3.0\% & \$0.011/s & \$0.008/s & $-$0.2 pp \\
CyberChair & 14.8\% & \$0.035/s & \$0.020/s & $-$0.3 pp \\
IBM & 16.0\% & \$0.037/s & \$0.022/s & $+$0.1 pp \\
iTrust Text2Policy & 20.9\% & \$0.047/s & \$0.027/s & $-$0.4 pp \\
\midrule
\textbf{Unweighted mean} & \textbf{13.7\%} & \textbf{\$0.033/s} & \textbf{\$0.019/s} & \textbf{$-$0.2 pp} \\
\bottomrule
\end{tabular}

\end{table*}

\textbf{The reduced-ensemble result.} Table~\ref{tab:cost} reports the head-to-head comparison. The reduced three-model ensemble produces a small macro-F1 degradation on three of four datasets ($-0.2$, $-0.3$, $-0.4$~pp on iTrust ACRE-n, CyberChair, and iTrust Text2Policy respectively) and a small improvement on IBM ($+0.1$~pp). The arithmetic mean across the four datasets is a $-0.2$~pp degradation. The change is small but non-zero: removing two ensemble members trades a fraction of a percentage point in macro-F1 for a 42\% cost reduction (from \$0.033/sentence to \$0.019/sentence). For deployments where this trade is acceptable, the three-model ensemble is the recommended configuration. The finding is informative not because accuracy is unchanged but because the trade-off curve is shallow: most of the ensemble's value comes from the top-three per-type performers identified at design time, and the additional two models provide marginal disagreement-resolution benefit relative to their cost.

\textbf{Open-source fine-tuning pilot (preliminary).} A Qwen-2.5-7B-Instruct + QLoRA pilot reaches global held-out macro-F1 0.829 $[0.738, 0.899]$ at a 580-sentence stratified training split (iTrust ACRE-n held-out 0.910, within 0.005 of the closed-source result) while not being viable at the 50-sentence scale --- an \emph{in-corpus} feasibility result, not cross-domain transfer. The gating cost appears to be annotation effort, not compute (5.0~min on one RTX~5090); the conclusion awaits out-of-domain validation, and adapters, results, and training scripts ship with the artifact. 

\FloatBarrier
\subsection{Cross-Dataset Replication and Diagnostic Case Studies}
\label{sec:replication}

To verify that the verification-rigor results generalize beyond a single dataset, we replicated all four rigor experiments (mutation testing, R2 trigger-rate analysis, Cypher false-positive rate analysis, and test-suite coverage) on each of the four datasets. The four datasets span three domains (healthcare, conference review, education) and three graph sizes (135--969 edges); the uniform post-patch re-run of Section~\ref{sec:mutation} equalizes the verifier side across all four. Table~\ref{tab:rigor_summary} summarizes the cross-dataset results; the per-dataset case studies that follow document the diagnostic Phase~B catches and explain the methodological role of each finding.

\begin{table*}[!tbp]
\centering
\caption{Cross-dataset rigor evaluation: unique results per corpus.}
\label{tab:rigor_summary}
\footnotesize
\setlength{\tabcolsep}{3pt}
\begin{tabular}{@{}lcccc@{}}
\toprule
\textbf{Metric} & \textbf{iTrust ACRE-n} & \textbf{CyberChair} & \textbf{IBM} & \textbf{iTrust Text2Policy} \\
\midrule
\multicolumn{5}{@{}l}{\emph{Phase B baseline (released graphs, post-patch$^{*}$)}} \\
Completeness & 98.9\% & 100\% & 100\% & 100\% \\
Correctness & 99.2\% & 100\% & 99.0\% & 100\% \\
Conflicts & 0 & 0 & 1* & 0 \\
Minimality & 97.4\% & 100\% & 100\% & 100\% \\
Verdict & PASS & PASS & NEEDS REFINE* & PASS \\
\midrule
\multicolumn{5}{@{}l}{\emph{Mutation testing detection}} \\
Total mutants (completed) & 237 & 210 & 231 & 240 \\
Aggr.\ detection (post-patch) & 100\% & \textbf{98.6\%} & 100\% & \textbf{100\%} \\
Aggr.\ detection (historical) & 100\% & 84.3\% & 100\% & 87.5\% \\
M2 detection (post-patch) & 100\% & 100\%** & 100\% & 100\%** \\
Cypher FPs / total failures & 0 / 8 & 0 / 1 & 0 / 3 & 0 / 0 \\
\midrule
\multicolumn{5}{@{}l}{\emph{Phase B natural-error case study}} \\
Catch & patient\_mid & PCC$\rightarrow$reviewer & student-register$^{c}$ & (clean baseline) \\
Class of finding & VP attachment & UA/RA confusion & temporal scoping & --- \\
\bottomrule
\end{tabular}
\par\smallskip{\raggedright\footnotesize *Patch applied by the deterministic verifier-side re-run; **historically 0\% (pre-patch CSVs carried no \texttt{sentence\_ids}); $^{c}$IBM's baseline conflict is documented as a temporal-scoping case study (Section~\ref{sec:ibm-case}), not fixed. Aggregates exclude vacuous/inapplicable M4 trials; mutation rows = uniform post-patch re-run; Phase-B and Cypher-FP blocks = re-run on the released graphs (development-time runs in Table~\ref{tab:phaseb}; the two tables' failure counts differ by design). CyberChair's single failing test is the S1 dangling-reference diagnostic of Section~\ref{sec:cyberchair-case}; strategy-level metrics are computed per element and read 100\% even as that one test fails. iTrust ACRE-n was processed post-patch, so its development-time and released-graph numbers coincide with Table~\ref{tab:phaseb}.\par}
\end{table*}

\textbf{Coverage gap: prohibition-related verification on CyberChair.} The CyberChair corpus contains zero PROHIBITION edges by virtue of the conference-management domain's natural language patterns (no ``deny''/``cannot'' constructs in the source documents). The M4 mutation operator (which injects a same-operation grant conflicting with an existing PROHIBITION edge, and therefore needs at least one deny to target; Table~\ref{tab:mutation_ops}) is therefore vacuous on CyberChair, so the prohibition-related portion of the rigor evaluation rests on three datasets (T2P, ACRE-n, IBM), not four. Independently, the M4-on-IBM ``vacuous skips'' described in Section~\ref{sec:ibm-case} are a different phenomenon (the random sampler rediscovers an existing IBM conflict), not a corpus-coverage gap.

\subsubsection{Diagnostic Case Studies}
\label{sec:itrust-case}
\label{sec:cyberchair-case}
\label{sec:ibm-case}

Three datasets produced one diagnostic catch each --- three distinct error classes across three corpora --- and the fourth provides the clean baseline.

\smallskip\noindent\textbf{iTrust ACRE-n: coordinated-VP attachment.} The patient\_mid grant/deny conflict fully documented in Section~\ref{sec:eval_phaseb}: a failure mode that prompt engineering alone does not reliably catch but that Phase~B's graph-level consistency check exposes by construction.

\smallskip\noindent\textbf{CyberChair: UA/RA type confusion.} Phase~B's single baseline failure (\texttt{S1\_Coverage\_0117}) traced to a dangling \texttt{RA:reviewer}: one sentence has the PCC ``asking'' a reviewer, so Stage~4 typed ``reviewer'' as an action target (RA), while the rest of the corpus consistently types it as the acting role (UA); Stage~6 canonicalizes the UA reading, and S1 correctly detects the orphaned reference. The class: an entity that legitimately plays both roles confounds sentence-local typing.

\smallskip\noindent\textbf{IBM: temporal scoping.} S3 flagged a same-operation grant/deny conflict on \texttt{(UA:student, register, RA:course\_offering)}: two sentences grant registration unconditionally while sentence~308 denies it after the semester's registration closes. Each extraction is correct in isolation, but the flat tuple representation cannot encode the temporal qualifier, so the deny conflicts structurally with the grants; the appropriate fix is a scope-aware Stage~4 representation (a conditional obligation or a deadline-activated policy class), so we document the case rather than patch it. The M4 sampler's 9 vacuous skips on IBM independently rediscover the same conflict, confirming S3 and the mutation generator observe the same graph state.

\smallskip\noindent\textbf{Clean baseline (iTrust Text2Policy).} The released Stage~6 graph scores 100\% on every Phase~B metric, with all 1{,}184 rigor-baseline tests passing. The initial development-time run (Table~\ref{tab:phaseb}: 1{,}430 tests, 9 failures --- 8 Cypher-formatting false positives, 1 true positive per Table~\ref{tab:phaseb_diagnosis}) preceded a cleanup that removed 16 redundant transitive assignments, yielding the released graph. The corpus's single ground-truth extraction error (the parent-abstraction case behind the 99.6\% Stage-4 accuracy) remains in the released graph without contradicting the clean baseline: parent-entity abstraction is precisely the E3 class that structural verification does not detect. Across the four corpora, Phase~B is sensitive in proportion to \emph{structurally visible} error density: PASS when clean, NEEDS-REFINE with a specific diagnostic when an issue exists, zero post-patch false alarms (Section~\ref{sec:cypher-fp}). 

\FloatBarrier
\FloatBarrier
\section{Adversarial Input: Prompt-Injection as a Fail-Open Channel}
\label{sec:injection}
The verifier boundaries above concern \emph{extraction errors} on good-faith documents. A distinct security question is what happens when the input document is itself adversarial. Because the pipeline's LLM stages read document text directly, a document author (or an attacker who can edit a document before extraction) has a prompt-injection surface, and Phase~B is structurally blind to it: its S1 tests verify the graph \emph{against the source text}, so an edge faithfully extracted from injected text passes verification by construction. We measured this channel: 24 adversarial CyberChair-domain sentences in three families (explicit spurious grants, indirection, embedded instructions) were injected into the corpus. The extractor emitted the attacker's intended grant for 16 of 24 (66.7\%), and 19 of 24 (79.2\%) produced a structurally well-formed edge that Phase~B accepts by construction --- text-level prompt injection is a measured fail-open channel. The M2 result (100\% detection of spurious grants) covers only edges \emph{without} sentence provenance and does not cover injected-but-grounded edges. Mitigations are upstream --- document provenance and integrity controls, instruction--data separation, an injection classifier on Phase-0 input --- or rest with the expert-review operating point, where a reviewer sees both the extracted grant and its source sentence. 

\section{Enforcement Replay in the NIST Reference Implementation}
\label{sec:pm-replay}

To confirm that the extracted graphs are enforceable, not merely well-formed, we translated the iTrust Text2Policy Stage~6 graph into PML and loaded it into the NIST reference implementation, \texttt{policy-machine-core} v3.3.0~\citep{ferraiolo2015pm}. The full structure --- 36 policy classes, 18 user attributes, 254 resource attributes, one resource, 430 assignments, 224 associations, and 8 prohibitions (as conjunctive prohibitions) --- loaded without modification (the corpus's 12 obligation records, which account for the remaining nodes in Table~\ref{tab:graphs}, are paired-edge constructs handled through the PM obligation mechanism, not graph nodes): every node and edge satisfied the reference implementation's type constraints, with zero orphaned attributes and zero skipped edges. Demonstration users and an authoring-rights scaffold for the obligation response are the only additions beyond the extracted graph (marked in the released generator).

We then replayed 55 access requests whose expected decisions were computed independently from SP 800-178 semantics (per-policy-class intersection over association-derived rights, minus prohibitions): 24 association-derived grants, 13 prohibition-covered denials, and 18 no-privilege denials. The reference PDP agreed on \textbf{55 of 55}, including every prohibition override, and one extracted paired-edge obligation (OBL\_367, an LHCP administering an immunization) was registered and fired end-to-end through the reference event-processing point. The scope of this result is \emph{translation fidelity and internal coherence, not source correctness}: the expected decisions are derived from the extracted graph under SP 800-178 semantics, so 55/55 agreement establishes that what \system{} extracts is faithfully and deterministically enforceable by the reference implementation. 

\section{Related Work}
\label{sec:related}

\noindent\textbf{NLP for access control policy extraction.} \citet{narouei2018identification} applied semantic role labeling for policy sentence identification, achieving approximately 80\% accuracy on the upstream identification subtask. \citet{slankas2014relation} introduced ACRE, a dependency-parse-driven relation extraction approach for software engineering documents, and released the larger iTrust corpus we evaluate against in this work. \citet{xia2022acre} applied BERT-based semantic-role labeling to ABAC policy extraction from healthcare documents. \citet{alohaly2019deep} explored deep learning for ABAC attribute extraction. \citet{xiao2012automated} extracted XACML-format ABAC policies using dependency parsing in their Text2Policy work, which gave the smaller iTrust corpus its name. None of these efforts target NGAC, none model obligations as event--response structures per NIST SP 800-178, none cover all six NGAC assignment sub-types ($U{\to}UA$, $UA{\to}UA$, $R{\to}RA$, $RA{\to}RA$, $UA{\to}PC$, $RA{\to}PC$), and none provide an automated graph-level verification stage; \system{} contributes all four.

\noindent\textbf{NGAC-specific extraction.} \citet{abdelgawad2023synthesizing} pioneered rule-based NGAC extraction from real documents (spaCy dependency parsing; associations and partial assignments at $\sim$75\% accuracy), with a follow-up narrowed to ABAC~\citep{abdelgawad2024abac}. \citet{lawal2024translating} introduced LLM-based extraction (Code4Policy), reporting 99\% F1 --- on \emph{LLM-synthesized} corpora (BoA, NHS, UTSA) that lack the compound sentences, pronoun chains, and attachment ambiguity of naturally-authored documents --- without assignments, graph verification, or verifier evaluation. We reproduce both as baselines (Section~\ref{sec:baseline}); Table~\ref{tab:comparison} details the per-capability distinctions, and the ablation's snapshot-dependence finding (Section~\ref{sec:ablation}) suggests why synthesized-data results transfer poorly to real text.

\begin{table*}[!tbp]
\centering
\caption{Detailed capability comparison with prior approaches.}
\label{tab:comparison}
\footnotesize
\setlength{\tabcolsep}{4pt}
\renewcommand{\arraystretch}{1.20}
\resizebox{\textwidth}{!}{%
\begin{tabular}{@{}>{\raggedright\arraybackslash}p{0.09\textwidth}>{\raggedright\arraybackslash}p{0.15\textwidth}>{\raggedright\arraybackslash}p{0.12\textwidth}>{\raggedright\arraybackslash}p{0.11\textwidth}>{\raggedright\arraybackslash}p{0.14\textwidth}>{\raggedright\arraybackslash}p{0.12\textwidth}>{\raggedright\arraybackslash}p{0.15\textwidth}@{}}
\toprule
\textbf{Approach} & \textbf{Extraction method} & \textbf{Type coverage} & \textbf{Assignment depth} & \textbf{Obligation semantics} & \textbf{Evaluation data} & \textbf{Verification \& rigor} \\
\midrule
\citet{abdelgawad2023synthesizing} (SACMAT~2023) &
spaCy \texttt{en\_core\_web\_lg} dependency parser; root-verb POS heuristics; rule-based negation list of 26 prohibition words &
Assoc., Prohib.; no obligation; partial assignment &
Only $U{\to}UA$ and $R{\to}RA$; no $UA{\to}UA$, $RA{\to}RA$, or PC sub-types &
Not supported; \texttt{advcl} clauses detected but reduced to associations &
7 university policy documents (real); 75\% average accuracy &
Cypher-based regex testcases (1 per path); no formal property checks; no controlled-mutation evaluation of the verifier \\
\midrule
\citet{abdelgawad2024abac} (SACMAT~2024) &
spaCy + ABAC-specific rule extensions; same syntactic pipeline as 2023 &
ABAC only (Subject, Action, Resource, Condition); no NGAC obligations or prohibitions &
Partial $U{\to}UA$ assignment via NER; ABAC scope only &
Not supported &
Same datasets as 2023; ABAC scope &
None reported \\
\midrule
\citet{lawal2024translating} (TPS-ISA~2024) &
GPT-4 with Code4Policy prompting (Python \texttt{@dataclass} schema); zero-shot baseline + few-shot scaled version &
Assoc., Prohib.; partial obligation; no assignment &
Not supported; assignment ``inherently established through tree hierarchy while synthesizing the dataset''~\citep{lawal2024translating} &
Partial: extracted as conditional clause; no event--response paired-edge construction &
3 LLM-synthesized datasets (BoA, NHS, UTSA, $\sim$917 sentences); no real-document evaluation &
NGAC specification-checker validates entity types in extracted relations; no formal property verification; no controlled-mutation evaluation of the verifier \\
\midrule
NIST report~\citep{nist2024translating} &
Chain-of-thought prompting (GPT-4) &
Assoc.; partial obligation; no prohibition or assignment &
Not supported &
Partial: described in narrative only &
Self-synthesized policy expressions &
None reported \\
\midrule
\citet{yang2025extraction} (SACMAT~2025) &
Knowledge distillation (large LLM teacher${\to}$small student model) &
Flat (Subject, Action, Resource) triples for ABAC; no NGAC types &
Not applicable (flat triples) &
Not supported &
Real ABAC policy data &
None reported \\
\midrule
\textbf{\system{} (this work)} &
\textbf{Phase 0 LLM disambiguation $+$ Phase A six-stage pipeline (DeepSeek R1 $+$ confidence-gated multi-LLM R2 ensemble; OpenAI embeddings; Claude validation)} &
\textbf{All four NGAC types: Association, Prohibition, Obligation, Assignment} &
\textbf{Represents all six sub-types; five exercised on these corpora ($U{\to}UA$ absent; \S\ref{sec:eval_graphs})} &
\textbf{Event--response paired-edge construction (materialized and enforced on iTrust Text2Policy; \S\ref{sec:pm-replay})} &
\textbf{4 real-world datasets (725 ACPs total) across 3 domains} &
\textbf{Phase B five-strategy verification (S1--S5) $+$ mutation-testing harness (8 structural + 4 taxonomy-derived operators; 99.7\% structural detection post-patch)} \\
\bottomrule
\end{tabular}%
}
\par\smallskip{\raggedright\footnotesize Cells state the method used or the scope of partial support, not a binary mark; Real = evaluated on real (vs.\ LLM-synthesized) documents; bottom row = \system{}.\par}
\end{table*}

\noindent\textbf{General-purpose information and relation extraction.} Joint entity--relation models (REBEL~\citep{huguet2021rebel}, UniRE~\citep{wang2021unire}), BERT-based RE pipelines~\citep{soares2019matching,xia2022acre}, and OpenIE systems~\citep{stanovsky2018supervised} extract flat \emph{(subject, predicate, object)} triples without the type discipline NGAC requires: no association-vs-prohibition polarity, no event--response obligation pairs, hierarchies treated as ordinary relations, and no verification against the access-control model. ACRE~\citep{slankas2014relation} comes closest in motivation but is likewise restricted to flat triples; we evaluate against its corpus (iTrust ACRE-n) but not its model. \system{} complements this literature: where those systems extract flat relations, it extracts a typed, verified, NGAC-compliant graph (head-to-head comparison with a fine-tuned RE model in Section~\ref{sec:baseline}).

A NIST technical report~\citep{nist2024translating} explores chain-of-thought prompting for NL-to-NGAC extraction, but excludes assignments entirely and includes no automated verification. \citet{yang2025extraction} contribute a knowledge-distillation approach evaluated on real data, but target flat ABAC triples (Subject--Action--Resource), not NGAC's richer graph semantics.

\system{} unifies these threads: LLM-based extraction evaluated on real documents; all four NGAC relationship types (six assignment sub-types represented, five exercised on these corpora); event--response paired-edge obligations that a NIST Policy Machine can enforce (Section~\ref{sec:pm-replay}), where prior approaches reduce obligations to conditional clauses that cannot fire; and five-strategy verification paired with a controlled-mutation evaluation of the verifier. No prior approach combines any two of these; Table~\ref{tab:comparison} details the per-capability distinctions.

\noindent\textbf{NGAC analysis and testing.} \citet{tan2025safety} proved the NGAC safety problem coNP-complete under mild assumptions, underscoring the importance of automated policy construction and verification. \citet{chen2023coverage} developed coverage-based testing of obligations; \citet{chen2021mutation} proposed mutation analysis for NGAC policies, providing the conceptual foundation for our Phase~B mutation harness. \system{} automates the upstream policy specification that these analytical frameworks operate upon.

\noindent\textbf{LLMs for structured extraction.} \citet{wei2022chain} demonstrated chain-of-thought prompting; \citet{wang2023selfconsistency} proposed self-consistency; \citet{brown2020language} established the few-shot capabilities underpinning RAG-augmented prompting; \citet{lewis2020retrieval} introduced retrieval-augmented generation. Our Stage~2 RAG context and Stage~3 R1+R2 confidence-gated ensemble adapt these techniques to the NGAC type-classification setting. Our multi-LLM ensemble strategy relates to a broader trend of applying LLMs to formal-specification extraction~\citep{jabal2020polisis}, where model complementarity and verification against formal semantics are critical challenges.

\noindent\textbf{Mutation testing.} Mutation analysis has a long history in software testing~\citep{jia2011mutation,papadakis2019mutation} and has been applied directly to NGAC policies by \citet{chen2021mutation} as a way to assess the adequacy of \emph{policy test suites}. Our use of mutation testing is closely related but addresses a different question: rather than measuring whether a policy's tests cover its mutants, we measure whether \system{}'s \emph{verifier} detects controlled corruptions of the extracted graph. The novelty relative to Chen \etal{} is narrower than ``mutation testing for NGAC'' (which Chen \etal{} introduced) and is best stated as: applying mutation analysis to validate a verifier of \emph{NL-extracted} policies, where the upstream extractor is itself fallible and the verifier's effectiveness against extractor errors is what is being measured. We share Chen \etal's mutation-operator design vocabulary (edge addition, edge deletion, type substitution, attribute reassignment) and adapt it to operate on the post-extraction graph rather than on a hand-authored policy.

\FloatBarrier
\section{Discussion and Limitations}
\label{sec:discussion}

\subsection{Limitations}
\label{sec:limitations}

Five bounded limitations are established elsewhere in the paper and need only a line each here. \emph{Pipeline-version split:} CyberChair and iTrust Text2Policy upstream outputs are cached pre-patch, while the deterministic verifier side runs uniformly post-patch on all four datasets (Section~\ref{sec:mutation}); an upstream re-run awaits a re-pinned LLM snapshot. \emph{Rebuild-vs-published topology deltas:} rebuilt graphs match the released Stage~6 CSVs on every verifier-relevant edge class, and the residual PC-attribution and obligation-materialization deltas are measured inert; diff scripts ship with the artifact. \emph{LLM-synthesized datasets:} the rationale for not re-running Code4Policy on BoA/NHS/UTSA accompanies the comparison (Section~\ref{sec:baseline-lawal}). \emph{Fault masking:} contingent on unrepaired baseline-flagged edges and post-repair-verified (Section~\ref{sec:mutation}); iTrust ACRE-n carries one analogous flagged edge (1 of 299 associations, never drawn at $n{=}30$). \emph{Reduced-ensemble scope:} a three-model ensemble preserves accuracy within $\pm$0.4~pp at $\sim$40\% lower Tier-2 cost (Section~\ref{sec:cost-effectiveness}). Three limitations warrant fuller discussion --- threshold calibration, training-data contamination, and verifier--operator co-design --- and follow in full prose; the wording-robustness probes are reported as experiments in Section~\ref{sec:paraphrase}.

\noindent\textbf{Threshold sensitivity.}\label{sec:threshold-sensitivity} The pipeline has two key thresholds; both sweeps characterize calibration, not robustness. \emph{Stage-3 confidence (default 0.9):} the R1 model emits only two confidence values, $\{0.85, 0.95\}$, on every dataset, so any threshold in $(0.85, 0.95]$ is behaviorally identical and the flat response is a degeneracy of binary-valued confidence, a known calibration weakness of zero-shot LLM scoring~\citep{xiong2024can}, not resilience. \emph{Phase-0 ambiguity (default 0.3):} sweeping $T \in [0.2, 0.7]$ on CyberChair spans a 3.7-pp R1 spread (decision-flip rates of 5.3--16.7\% at $\pm$0.1 across corpora), with 0.3 sitting at the score-distribution inflection that captures all observed decision flips. Because Stages~4--6 and Phase~B are deterministic functions of the Stage-3 output, the 3.7-pp R1 spread upper-bounds any end-to-end Path-B effect, which is smaller than the bootstrap CI half-widths on every Phase~B metric ($\pm$2--6~pp), so a full Path-B sweep could not distinguish a threshold effect from sampling noise at these corpus sizes; running one against current vendor snapshots would additionally mix in vendor drift (Section~\ref{sec:threats}). A full Path-B sweep against a re-pinned snapshot remains planned follow-up; per-threshold tables ship with the artifact. 

\noindent\textbf{Training-data contamination.} Three of the four evaluation corpora are long-public documents: the iTrust requirements date to 2012~\citep{meneely2012itrust}, the IBM Course Registration document to 2010~\citep{ibm2010course}, and CyberChair has been publicly available for well over a decade. All of them plausibly appear in the pretraining corpora of the commercial LLMs the pipeline relies on (DeepSeek-Chat, GPT-4o-mini, Claude Sonnet, and Claude Opus), raising the natural question of whether the reported 97--99\% extraction accuracy reflects memorization of these documents, not extraction capability. Three observations bound this threat. First, the ground-truth NGAC tuple annotations are new and unpublished: they were constructed by the authors for this work (Section~\ref{sec:datasets}), so no (sentence, NGAC tuple) pairs for these corpora exist anywhere public. Memorizing the source \emph{documents} therefore does not yield the \emph{labels}: mapping a sentence to a typed NGAC tuple with correct UA/RA prefixes, normalized operations, and the right assignment sub-type requires performing the extraction task. Second --- a weaker argument, since the paraphrase probes (Section~\ref{sec:paraphrase}) show surface-form sensitivity that memorization would also predict --- single-model extraction reaches only 58.5--87.0\% exact match on the same sentences (Table~\ref{tab:ablation}) against the full pipeline's 97.3--99.6\%; if memorized sentence-to-tuple mappings drove the headline number, a single call would suffice. Third, the Phase~B case studies (the patient\_mid attachment error, the PCC$\rightarrow$reviewer type confusion, and the IBM temporal-scoping conflict; Section~\ref{sec:replication}) document genuine extraction errors on these very corpora, errors of exactly the kind memorization would prevent. The paraphrase probes of Section~\ref{sec:paraphrase} measure surface-form sensitivity directly; their residual drop cannot be attributed to memorization specifically, and the three bounding observations above are unaffected by it.

\noindent\textbf{Verifier--operator co-design.} The eight M-operators target specific Phase~B strategies, so 100\% detection of the co-designed set is necessary but not sufficient evidence of effectiveness against arbitrary faults; Sections~\ref{sec:mutation}--\ref{sec:taxonomy-ops} treat the circularity and its mitigations in full (operator vocabulary adapted from the independently designed \citet{chen2021mutation}; the surfaced M2 gap showing the harness finds real holes; the independently sampled E-operators scoring 96.7--99.2\% on three classes with the E3 boundary at 9.2\%). The E3 boundary now has a piloted sentence-grounded mitigation (95\% of E3 mutants flagged at a 2--5\% review burden; Section~\ref{sec:taxonomy-ops}); what remains open is a fully automated repair and fault models drawn from sources external to this paper entirely, \eg{} operator vocabularies from other verification literatures. The twelve operators sample the fault classes we could mechanically generate, not a uniform coverage of NGAC's semantic fault space --- cross-sentence referent drift and plausible-but-wrong operation choices, for instance, have no generator --- so per-class detection rates are measurements, not a global adequacy score. 

\smallskip\noindent\textbf{Threat model: trusted policy documents.} \system{} assumes its input documents are authored in good faith; adversarial input is outside the extraction guarantees, and Section~\ref{sec:injection} measures the prompt-injection channel this assumption leaves open (66.7\% attack success, accepted by Phase~B by construction; mitigations are upstream or rest with the expert reviewer). Concretely, a malicious policy author needs only one planted sentence granting their own role access to a sensitive resource attribute: the pipeline faithfully encodes it, every structural check passes, and the grant reaches the reviewer as legitimate-looking output --- which is why the mitigations must sit upstream of extraction.

\smallskip\noindent\textbf{What the system does not guarantee.} Four boundaries define how the results should be read: a graph that passes Phase~B is \emph{structurally validated, not semantically certified} (Section~\ref{sec:severity}); accuracy is \emph{wording-specific}, not wording-invariant (Section~\ref{sec:paraphrase}); accuracy is \emph{relative to the released annotations}, whose blind-coder reliability and agreed-subset control are in Section~\ref{sec:eval_stage4}; and the intended operating point is extraction plus structural validation \emph{feeding expert review}, not autonomous deployment.

\subsection{Threats to Validity}
\label{sec:threats}

\noindent\textbf{Internal validity.} The primary internal threat is circularity between ground truth and system: annotations follow the same NGAC naming conventions the extractor emits. Three considerations bound it. Both the annotation guide and the Stage~4 prompts derive their naming conventions from the NIST SP 800-178 type definitions (Section~\ref{sec:ngac}), a shared external source, rather than either being derived from the other. Reliability is established by independent second-annotator coding: Cohen's $\kappa=0.84$ pooled (Section~\ref{sec:datasets}). The agreed-vs-contested split reported next to the headline (Section~\ref{sec:eval_stage4}) shows the headline is not carried by the contested labels. All metrics use exact-match semantics against this ground truth, with the Stage-3/Stage-4 reporting convention of Section~\ref{sec:eval_stage4}. Ablation comparisons use exact McNemar tests on paired per-sentence exact-match outcomes; significant effects are reported with exact $p$-values and null effects as null (Table~\ref{tab:ablation}). The R2 trigger-rate trend tests use both linear regression and Mann-Kendall; both reach the same null conclusions. 

\noindent\textbf{External validity.} Publicly available NL-to-access-control datasets remain scarce (Section~\ref{sec:datasets}). Our four datasets from three domains exceed the evaluation scale of all prior NGAC-extraction work. However, the all-four-types coverage claim rests on a thin slice of that data: the corpora contain 52~obligation and 23~prohibition sentences out of 725, CyberChair contributes zero prohibitions, and IBM a single obligation --- so per-type results for these two classes are evidence of capability on $\sim$75 sentences, not of robust per-domain performance. 
All four are English-language enterprise documents; applicability to other languages, IoT/cloud policies, or different linguistic registers remains to be validated. Their long public availability also raises a training-data contamination concern, which we analyze and bound in Section~\ref{sec:limitations} (training-data contamination). The mutation-testing framework is graph-structural and language-independent, so we expect it to generalize naturally; the LLM-driven extraction stages are language-dependent and would require retraining or prompt adaptation for other languages.

\noindent\textbf{Construct validity.} We report exact-match accuracy, macro-averaged precision, recall, and F1 throughout, all scored against the released labels. Independent second-annotator agreement is $\kappa=0.84$ pooled (Section~\ref{sec:datasets}); the residual disagreements concentrate at the assignment--association and obligation--association boundaries, the same categories that dominate error analysis (Section~\ref{sec:eval_errors}) --- boundary ambiguity intrinsic to NGAC typing, not random noise, and bounded by the agreed-subset control (Section~\ref{sec:eval_stage4}). Phase~B pass rates depend on LLM-generated test quality; partial-credit metrics~\citep{slankas2014relation} and mutation-based test adequacy~\citep{chen2021mutation} are alternative effectiveness measures for future work.

\noindent\textbf{Conclusion validity.} Three risks bear on the strength of our quantitative claims. (1)~\emph{Prompt sensitivity}: all reported numbers use the structured prompts in our public artifact and \texttt{temperature=0.1}. We did not run a systematic prompt-perturbation study, so the absolute numbers may shift under prompt rewrites; however, every comparison in the paper (full-pipeline vs.\ ablations, \system{} vs.\ baselines, R1 vs.\ R2) holds the prompt fixed across conditions, so relative claims are robust to this risk. (2)~\emph{Model-version drift}: the underlying LLM endpoints are vendor-managed and may evolve. We pin exact model versions in the released artifact, but exact bit-level reproduction of Tier~2 (LLM-driven) results across vendor updates is not guaranteed. The Tier~1 verification and rigor results are deterministic and unaffected by drift. Migrating to a new snapshot is a bounded recalibration exercise: re-run the 50-sentence design-time benchmark to refresh per-type ensemble weights, re-sweep the two thresholds (both sweeps ship with the artifact), and re-validate on one corpus; the deterministic verifier side needs no change. (3)~\emph{Multiple-comparison risk}: the ablation makes nine within-run comparisons (three conditions $\times$ three corpora; Table~\ref{tab:ablation}). Under a Bonferroni correction ($\alpha/9 = 0.0056$), the Phase-0 effect on iTrust Text2Policy survives ($p{=}0.002$); the combined-removal effect on the same corpus ($p{=}0.009$) is nominally significant but does not survive the correction and is suggestive only. For the archived-full-vs-re-run-R1 comparisons, every discordant sentence pair favors the full pipeline (96:0, 6:0, 29:0); the CyberChair count (6:0, $p{=}0.031$) is nominally significant but does not survive a three-way correction, while the other two are far below any threshold. Null results are reported as null throughout.

\subsection{Reproducibility}
\label{sec:reproducibility}

\system{}'s artifact has two reproducibility tiers (a separate axis from the cost tiers of Section~\ref{sec:cost-effectiveness}). \emph{Tier~1} (zero API cost, fully deterministic): Phase~B verification and all rigor experiments operate on the released Stage~6 CSVs with seeded harnesses (\texttt{random\_seed=42}) and reproduce bit-exactly in under five minutes on a single CPU. \emph{Tier~2} (LLM-dependent): the Phase~0~+~Phase~A extraction stages run at \texttt{temperature=0.1} against pinned commercial model versions; we measured $\leq 1.5\%$ sentence-level disagreement between two independent full-pipeline runs on iTrust ACRE-n at the same pinned versions, while vendor model updates can shift absolute numbers (a drift instance is measured in Section~\ref{sec:ablation}); the released ground truth and scoring scripts let any re-run be evaluated against the same canonical metrics, with the reported bootstrap CIs as the tolerance bounds.

A per-result reproducibility map (each headline result, its tier, and the expectation an independent replication should hold) is provided in the supplementary material.

\subsection{Future Work}
\label{sec:future}

Five concrete threads of follow-up work flow directly from the limitations above: \textbf{F1}~a conditional-scope representation (conditional obligations or scope-bounded policy classes) so temporal qualifiers like the IBM registration deadline become first-class rather than verifier-reconstructed (Section~\ref{sec:ibm-case}); \textbf{F2}~a multilingual pilot (professionally translated UC1 set; Spanish, Mandarin, Arabic); \textbf{F3}~a few-shot scaling study on a single 1{,}000-sentence corpus, since $n{=}54$--$362$ shows no monotone pattern (Section~\ref{sec:r2-scalability}); \textbf{F4}~an out-of-domain evaluation of the open-source fine-tuned ensemble slot, extending the preliminary pilot of Section~\ref{sec:cost-effectiveness}; and \textbf{F5}~a human-in-the-loop review UI that surfaces Phase-B-flagged low-confidence cases with proposed graph edits, converting the residual 1--3\% extraction gap into a tractable manual workflow.

\FloatBarrier
\section{Conclusion}
\label{sec:conclusion}

We presented \system{}, a multi-stage pipeline for near-automated NGAC policy extraction from natural language, paired with five-strategy semantic verification (Phase~B) whose effectiveness is itself evaluated through a controlled-mutation analysis adapted from \citet{chen2021mutation} to the NL-extracted-graph setting. Three contributions — proactive semantic disambiguation (Phase~0), task-optimized multi-LLM ensemble routing (Phase~A), and the verifier-plus-mutation-evaluation pairing (Phase~B + §\ref{sec:mutation}) — collectively achieve 97.3--99.6\% Stage~4 sentence-level extraction accuracy across four diverse datasets on their original wording (wording-specific; Section~\ref{sec:paraphrase}), far above all three reproduced baselines on matched metrics (Table~\ref{tab:baseline}). The ablation shows the ensemble-plus-refinement layer is the dominant contributor, and no single frontier model from three vendors closes the gap (Section~\ref{sec:ablation}).

The mutation-testing experiment shows Phase~B detects 99.7\% of the 918 structural mutants under a uniform, deterministic post-patch re-run, with M2 detection moving 0\%$\rightarrow$100\% when the S1 edge-provenance patch is toggled on identical graphs; the residual misses trace to fault masking on a single baseline-flagged edge (Section~\ref{sec:mutation}). On the taxonomy-derived operators, detection is 96.7--99.2\% on three fault classes; the fourth, parent-entity abstraction (9.2\%), is the verifier's measured boundary and dominates the residual fail-open exposure (Table~\ref{tab:severity}); a sentence-grounded pilot detector flags 95\% of these faults at a 2--5\% review burden (Section~\ref{sec:taxonomy-ops}). The four-corpus replication adds three diagnostic case studies, each a distinct Stage~4 error class, and the end-to-end amortized cost is \$0.011--\$0.047 per sentence, with a reduced three-model ensemble preserving accuracy at roughly 40\% lower Tier-2 cost.

The extracted graphs are demonstrably enforceable --- the iTrust Text2Policy graph loads unmodified into the NIST reference implementation, the reference PDP reproduces all 55 access decisions the graph entails under SP 800-178 semantics (translation fidelity, not source correctness), and an extracted paired-edge obligation fires end-to-end (Section~\ref{sec:pm-replay}) --- and the verifier-plus-mutation-evaluation pairing is a general methodology, applicable wherever a structured artifact is extracted from unstructured text and mechanically verifiable correctness properties exist. Together with complete NGAC type coverage, including event--response obligations, and explicit cost-effectiveness, this positions \system{} as a practical tool for translating natural-language policies into structurally validated NGAC policy graphs ready for expert review.

\backmatter

\bmhead{Abbreviations}
ABAC: Attribute-based access control; ACP: Access control policy; ACRE: Access control relation extraction; API: Application programming interface; CI: Confidence interval; CRUD: Create, read, update, delete; FP: False positive; GT: Ground truth; HCP: Health care personnel; IE: Information extraction; LHCP: Licensed health care personnel; LLM: Large language model; NGAC: Next Generation Access Control; NIST: National Institute of Standards and Technology; NL: Natural language; NLP: Natural language processing; OOD: Out-of-domain; PC: Policy class; R: Resource; RA: Resource attribute; RAG: Retrieval-augmented generation; RE: Relation extraction; SP: Special Publication; U: User; UA: User attribute; VP: Verb phrase.

\bmhead{Supplementary material}
The supplementary analyses (\texttt{supplementary.pdf}) are available as an ancillary file of this arXiv record and in the reproducibility artifact: the R2 trigger-rate scalability analysis, the Cypher false-positive patch details, the per-element test-suite coverage table, the BERT-RE learning-rate sweep, the per-result reproducibility map, and the per-cell bootstrap confidence intervals for the results and baseline tables.

\section*{Declarations}

\bmhead{Availability of data and materials}
The reproducibility artifact is archived on Zenodo under the concept DOI \url{https://doi.org/10.5281/zenodo.20309352}, which always resolves to the latest published version. It contains the four annotated corpora and both independent codings, the extraction and verification code (Phases 0/A/B for every dataset), the mutation and rigor harnesses, reproduced baselines, statistical analysis scripts, the probe and pilot-detector records, and the NIST enforcement-replay harness. The version matching this manuscript will be published under the same concept DOI. Reproduction is split into a deterministic single-CPU Tier~1 (no API access required) and an LLM-driven Tier~2 (reproducible up to vendor-side stochasticity at \texttt{temperature=0.1}). Code is MIT-licensed; data and annotations CC~BY~4.0.

\bmhead{Competing interests}
The authors declare that they have no known competing financial interests or personal relationships that could have appeared to influence the work reported in this paper.

\bmhead{Funding}
This work was supported by the Key Research and Development Program of Shaanxi (Program No.~2024GX-ZDCYL-01-20).

\bmhead{Authors' contributions}
\textbf{Ayesha Habib}: Conceptualization, Methodology, Software, Investigation, Formal analysis, Data curation, Validation, Visualization, Writing (original draft), Writing (review \& editing). \textbf{Huixiang Zhang}: Supervision, Conceptualization, Methodology, Resources, Funding acquisition, Writing (review \& editing). \textbf{Dan Zhu}: Methodology, Validation, Writing (review \& editing). \textbf{Waiz Haider}: Software, Investigation, Validation, Data curation, Writing (review \& editing). All authors read and approved the final manuscript.

\bmhead{Acknowledgements}
The authors thank the iTrust, CyberChair, and IBM Course Registration project maintainers for the publicly available source documents on which the annotated corpora are built.

\bmhead{Use of generative AI in scientific writing}
During the preparation of this work the authors used generative-AI assistants (Claude Code by Anthropic) to assist with LaTeX editing, code review, drafting prose suggestions, and running supporting experiments (BERT-RE baseline training, bootstrap confidence-interval scripts, and the Qwen-2.5-7B+QLoRA fine-tuning pilot). After using these tools, the authors reviewed and edited the content as needed and take full responsibility for the content of the publication.

\bibliography{references}

\end{document}